# Efficient High-Quality Clustering for Large Bipartite Graphs

Renchi Yang
Hong Kong Baptist University
Hong Kong SAR, China
renchi@hkbu.edu.hk

Jieming Shi*
Hong Kong Polytechnic University
Hong Kong SAR, China
jieming.shi@polyu.edu.hk

## ABSTRACT

A bipartite graph contains inter-set edges between two disjoint vertex sets, and is widely used to model real-world data, such as user-item purchase records, author-article publications, and biological interactions between drugs and proteins. $k$-Bipartite Graph Clustering ($k$-BGC) is to partition the target vertex set in a bipartite graph into $k$ disjoint clusters. The clustering quality is important to the utility of $k$-BGC in various applications like social network analysis, recommendation systems, text mining, and bioinformatics, to name a few. Existing approaches to $k$-BGC either output clustering results with compromised quality due to inadequate exploitation of high-order information between vertices, or fail to handle sizable bipartite graphs with billions of edges.

Motivated by this, this paper presents two efficient $k$-BGC solutions, HOPE and HOPE+, which achieve state-of-the-art performance on large-scale bipartite graphs. HOPE obtains high scalability and effectiveness through a new $k$-BGC problem formulation based on the novel notion of high-order perspective (HOP) vectors and an efficient technique for low-rank approximation of HOP vectors. HOPE+ further elevates the $k$-BGC performance to another level with a judicious problem transformation and a highly efficient two-stage optimization framework. Two variants, HOPE+ (FNEM) and HOPE+ (SNEM) are designed when either the Frobenius norm or spectral norm is applied in the transformation. Extensive experiments, comparing HOPE and HOPE+ against 13 competitors on 10 real-world datasets, exhibit that our solutions, especially HOPE+, are superior to existing methods in terms of result quality, while being up to orders of magnitude faster. On the largest dataset MAG with 1.1 billion edges, HOPE+ is able to produce clusters with the highest clustering accuracy within 31 minutes, which is unmatched by any existing solution for $k$-BGC.

## 1 INTRODUCTION

A bipartite graph $\mathcal{G}$ contains two disjoint vertex sets, $\mathcal{U}$ and $\mathcal{V}$, with only inter-set edges. Bipartite graphs are omnipresent in the real world to model the purchase/view records between customers and products, the publication relationships between authors and articles, the biological interactions between drugs and protein complexes, and the ecological relationships between plants and pollinators.

A fundamental task in data mining is $k$-*Bipartite Graph Clustering* (hereafter $k$-BGC) which aims to group the vertices in the target vertex set, e.g., $\mathcal{U}$, into $k$ disjoint clusters, each of which is tightly-knit, based on the interplay between the two types of vertices $\mathcal{U}$ and $\mathcal{V}$ in the bipartite graph $\mathcal{G}$ (the clustering on the other set $\mathcal{V}$ naturally follows). $k$-BGC finds prevalent use in various fields, including social network analysis, recommendation systems, text mining, bioinformatics, and so forth. Practical applications include clustering customers based on customers' purchasing patterns [25, 59], categorizing documents based on document-word associations [11, 12], detecting groups of genes suitable for drug repurposing in gene-drug networks [34], and many others [10, 14, 16, 28, 38, 46, 66].

One simple treatment for $k$-BGC is to regard a bipartite graph as a unipartite graph, and then apply canonical graph clustering methods, e.g., [20, 39, 55], to get clustering results. However, this brute-force approach overlooks the unique topological properties of bipartite graphs with two disjoint vertex sets, and thus, yields sub-par clustering quality.

As reviewed in Section 6, there exists a plethora of studies with dedicated techniques developed particularly for bipartite graph clustering under various settings [8] (see references therein). In particular, one popular methodology, projection-based methods [23, 36, 52], first projects the input bipartite graph $\mathcal{G}$ into a unipartite graph containing the same-typed vertices via the one-mode projection [70, 71], and then applies conventional graph clustering methods over the projected graph for vertex clustering. These methods often generate extremely dense projected graphs with $O(|\mathcal{U}|^2)$ or $O(|\mathcal{V}|^2)$ edges in the worst case [62], as shown in Figure 1(a) with a extremely dense projected graph created due to the connections from all users to the phone, rendering them inefficient. Another category of techniques extends spectral clustering [12, 31] or statistical models (e.g., stochastic block models) [32, 67] to simultaneously group vertices of both types in bipartite graphs. To our knowledge, none of these methods effectively exploit the *high-order affinities* between vertices (vertices can reach each other through multiple hops along the edges), which intuitively is critical for deriving high-quality clustering results and has been shown effective in unipartite graph clustering [43, 65]. Consequently, most of them compromise cluster quality. To illustrate, we consider the customer-product purchase network in Figure 1(b). Suppose that there are two clusters to discover, i.e., $k = 2$. These methods may assign user D to the r.h.s cluster {E, F, G}, since doing so leads to a good graph partition with solely one cross-cluster edge (in purple), and they cannot discern the strong high-order affinity between users D and A, B, C. But, intuitively, user D is more likely to belong to the l.h.s cluster {A, B, C}, given that user D has as many as nine 3-hop paths to the basketball, football, and game console as users A, B, and C do, reflecting their similar tastes to the products. By contrast, user D possesses only one 3-hop path to either book or skirt and zero 3-hop path to stilettos purchased by users E, F and G, which indicates that they have utterly divergent shopping preferences and should be grouped into disparate clusters.

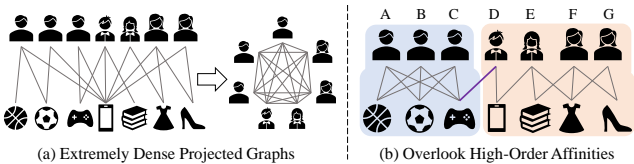

(a) Extremely Dense Projected Graphs     (b) Overlook High-Order Affinities

**Figure 1: Limitations of Existing Approaches**

On top of that, existing solutions incur prohibitive computational overheads, and are hard to scale to large-scale bipartite graphs with millions of nodes and billions of edges. For instance, on an MAG bipartite graph with 1.1 billion edges, almost all competitors fail to finish clustering within 2 days in experiments. An existing strong approach, BISBM-MCMC [67] takes over one week to process MAG. Summing up, existing $k$-BGC methods suffer from sub-optimal clustering quality, entail tremendous computational overheads and largely fail over massive bipartite graphs, or both.

To tackle the above challenges, we propose HOPE and HOPE+ that achieve superior performance for $k$-BGC, via a series of novel algorithmic designs and theoretical analysis over newly designed notion High-Order Perspective VEctors. Given a bipartite graph $\mathcal{G}$ with the target vertex set $\mathcal{U}$ to be clustered, it is crucial but challenging to capture the rich topological features for effective clustering. To this end, we first construct a weighted projected graph (WPG) $\mathcal{G}_{\mathcal{V}}$ over the other vertex set $\mathcal{V}$ of $\mathcal{G}$ with a new weighting scheme. Intuitively, if two vertices $u_i$ and $u_j$ in $\mathcal{U}$ have different (resp. similar) perspectives to WPG $\mathcal{G}_{\mathcal{V}}$, they tend to be in different clusters (resp. the same cluster). To realize this, we define a high-order perspective (HOP) per vertex $u_i \in \mathcal{U}$ via a random walk model from $u_i$ towards the WPG $\mathcal{G}_{\mathcal{V}}$, so as to quantify the personalized high-order semantics from $u_i$ to all vertices $\mathcal{V}$ in $\mathcal{G}_{\mathcal{V}}$. Based thereon, we formulate a new $k$-BGC clustering objective defined using the HOP vectors of vertices in $\mathcal{U}$.

The direct materialization of all HOP vectors for $\mathcal{U}$ is rather costly, as they constitute a dense HOP matrix $\mathbf{H} \in \mathbb{R}^{|\mathcal{U}| \times |\mathcal{V}|}$ whose calculation involves an infinite series of matrix multiplications with $O(|\mathcal{U}| \times |\mathcal{V}|^2)$ time. To mitigate the issue, we first present a fast method HOPE that builds up a low-rank approximation of $\mathbf{H}$ with theoretical accuracy guarantees, without explicitly materializing $\mathbf{H}$, and subsequently conducts clustering using the low-rank approximation as per our proposed objective. HOPE can achieve improved result quality, while being fast, compared with existing methods.

Furthermore, we propose HOPE+ that surpasses HOPE and existing methods in terms of both efficiency and effectiveness. Under the hood, through rigorous theoretical analysis, HOPE+ converts our HOP-based $k$-BGC objective into two optimization problems to be solved via our two-stage algorithmic framework. The first problem is optimized by a certain relaxed intermediate result obtained by an efficient partial eigen-decomposition; then in the second optimization, HOPE+ derives the final cluster result by minimizing its difference to the intermediate result via two rounding algorithms, namely FNEM and SNEM, when either Frobenius norm or spectral norm is applied to quantify the difference. The complexities of HOPE, HOPE+ (FNEM), and HOPE+ (SNEM) are all linear to the size of the input bipartite graph.

We empirically evaluate HOPE and HOPE+ against 13 competitors on 10 real datasets with ground-truth cluster labels. Extensive experiments establish that HOPE and HOPE+ obtain superior clustering quality on most datasets while being up to orders of magnitude faster. Particularly, on the billion-edge dataset MAG, HOPE+ with SNEM achieves a 22.1% conspicuous improvement in clustering accuracy and at least 9.5× speedup over existing methods.

To summarize, our contributions in this paper are as follows:
- We formulate an effective $k$-BGC objective based on the new High-Order Perspective (HOP) vectors that preserve vertex-specific high-order information over a weighted projected graph.
- We present HOPE that efficiently solves the objective via low-rank approximation, without materializing all HOP vectors, to output high-quality clustering results.
- We further develop HOPE+ that transforms the $k$-BGC objective into two optimization problems that are solved by efficient technical designs. We present two variants HOPE+ (FNEM) and HOPE+ (SNEM), when Frobenius norm or spectral norm is applied.
- Extensive experiments on 10 bipartite graphs validate the superiority of our methods to efficiently obtain high-quality clusters.

## 2 PROBLEM FORMULATION

### 2.1 Notations

Let $\mathcal{G} = (\mathcal{U} \cup \mathcal{V}, \mathcal{E})$ be a weighted bipartite graph, where $\mathcal{U}$ and $\mathcal{V}$ are two disjoint vertex sets, and the edge set is $\mathcal{E} = \{(u_i, v_j) \mid u_i \in \mathcal{U}, v_j \in \mathcal{V}\}$. Every edge $(u_i, v_j) \in \mathcal{E}$ is associated with a non-negative weight $w(u_i, v_j)$. The set of neighbors (*i.e.*, adjacent vertices) of a vertex $u_i \in \mathcal{U}$ (resp. $v_j \in \mathcal{V}$) is denoted by $\mathcal{N}(u_i) = \{v_j | (u_i, v_j) \in \mathcal{E}\}$ (resp. $\mathcal{N}(v_j) = \{u_i | (u_i, v_j) \in \mathcal{E}\}$).

In this paper, matrices are denoted by bold uppercase letters, *e.g.*, $\mathbf{M} \in \mathbb{R}^{n \times m}$ with $n$ rows and $m$ columns. The $i$-th row (resp. $j$-th column) of $\mathbf{M}$ is a length-$m$ (resp. length-$n$) vector, written as $\mathbf{M}_i$ (resp. $\mathbf{M}_{*,j}$), and the entry at the $i$-th row and $j$-th column of $\mathbf{M}$ is denoted by $\mathbf{M}_{i,j}$. The Frobenius norm of matrix $\mathbf{M}$ is $\|\mathbf{M}\|_F = \sqrt{\sum_{i=1}^{n} \sum_{j=1}^{m} |\mathbf{M}_{i,j}|^2}$, and the $L_2$ norm of vector $\mathbf{M}_i$ is $\|\mathbf{M}_i\|_2 = \sqrt{\sum_{j=1}^{m} |\mathbf{M}_{i,j}|^2}$. The trace of a square matrix $\mathbf{M} \in \mathbb{R}^{m \times m}$ is $\mathrm{Tr}(\mathbf{M}) = \sum_{i=1}^{m} \mathbf{M}_{i,i}$. We use $\mathbf{M}^\top$ to represent the transpose of $\mathbf{M}$ and $\mathbf{I}$ to represent the identity matrix with dimension implied by the context. A matrix $\mathbf{M}$ is said to have orthogonal columns (resp. rows) if it satisfies $\mathbf{M}^\top \mathbf{M} = \mathbf{I}$ (resp. $\mathbf{M}\mathbf{M}^\top = \mathbf{I}$). Table 1 lists the frequently used notations throughout this paper.

### 2.2 $k$-BGC

Given a weighted bipartite graph $\mathcal{G} = (\mathcal{U} \cup \mathcal{V}, \mathcal{E})$, the target vertex set to cluster (either $\mathcal{U}$ or $\mathcal{V}$), and the number $k$ of clusters, the goal of $k$-*Bipartite Graph Clustering* ($k$-BGC) is to partition the target vertex set into $k$ disjoint vertex clusters, $C_1, C_2, ..., C_k$, such that the vertices in the same cluster are closely connected in $\mathcal{G}$ via direct or indirect connections, whereas the vertices across clusters are distant from each other [23, 32, 67]. By default, we regard $\mathcal{U}$ as the target vertex set to cluster. It naturally follows when $\mathcal{V}$ is the target. Choosing $k$ is not a focus of this paper, and can be done by an additional step [7].

A critical challenge for $k$-BGC is to leverage the rich semantics hidden in the bipartite graph topology to obtain high-quality features of vertices for effective clustering. In the following subsections, we propose a new way to formulate the $k$-BGC objective. In a nutshell, we first construct a *weighted projected graph* (WPG) $\mathcal{G}_{\mathcal{V}}$ built over the counterparty $\mathcal{V}$ in Section 2.3, and then, for every vertex $u_i$ in $\mathcal{U}$, we define a *high-order perspective* (HOP) vector (Section 2.4), which captures the personalized multi-hop information from $u_i$ over the WPG $\mathcal{G}_{\mathcal{V}}$, and then formulate our $k$-BGC objective based on the HOP vectors of $\mathcal{U}$ (Section 2.5).

**Table 1: Frequently used symbols.**

| Symbol | Description |
|---|---|
| $\mathcal{G} = (\mathcal{U} \cup \mathcal{V}, \mathcal{E})$ | A bipartite graph of $\mathcal{G}$ with vertex sets $\mathcal{U}, \mathcal{V}$ and edge set $\mathcal{E}$. |
| $\mathcal{N}(u_i)$ | The neighbor set of $u_i \in \mathcal{U}$ in $\mathcal{G}$, $\mathcal{N}(u_i) \subseteq \mathcal{V}$ |
| $w(u_i, v_j)$ | The weight of edge $(u_i, v_j)$ in $\mathcal{G}$ |
| $\mathbf{P}$ | The transition matrix of $\mathcal{G}$ where $\mathbf{P}_{i,j} = p(u_i, v_j)$ in Eq. (1). |
| $\mathcal{G}_{\mathcal{V}}$ | The weighted projected graph (WPG) on $\mathcal{V}$ |
| $w_{\mathcal{V}}(v_j, v_l)$ | The weight of edge $(v_j, v_l)$ in $\mathcal{G}_{\mathcal{V}}$ |
| $\mathbf{Q}$ | The matrix of $\mathcal{G}$ where $\mathbf{Q}_{j,i} = \sqrt{p(v_j, u_i) \cdot p(u_i, v_j)}$. |
| $\mathbf{H}$ | The HOP vectors defined in Eq. (6). |
| $\mathbf{X}$ | The low-rank approximation of $\mathbf{H}$. |
| $\mathbf{C}$ | The VCMI matrix defined in Eq. (10). |
| $k$ | The number of clusters. |
| $\beta$ | The dimensionality of low-rank approximation. |
| $\alpha$ | The random walk decay factor. |

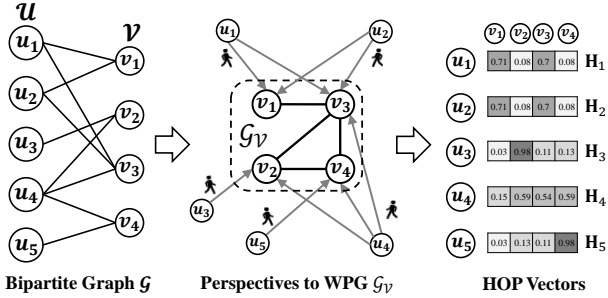

**Figure 2: Construction of HOP Vectors**

Figure 2 exemplifies the generation of HOP vectors. In the first place, we construct the WPG $\mathcal{G}_{\mathcal{V}}$ for all vertices in $\mathcal{V}$ on the basis of the topology surrounding each $v_j$ in $\mathcal{V}$ of the input bipartite graph $\mathcal{G}$. Thereafter, each vertex $u_i$ in $\mathcal{U}$ is represented by an HOP vector computed as per a proposed random walk model over WPG $\mathcal{G}_{\mathcal{V}}$ from $u_i$'s perspective. Intuitively, two vertices (*e.g.*, $u_1$ and $u_2$ in Figure 2) with similar perspectives towards all vertices in the WPG $\mathcal{G}_{\mathcal{V}}$ tend to have similar HOP vectors, while $u_1$ and $u_3$ are with relatively different HOP vectors since their perspectives (*i.e.*, connections) towards WPG $\mathcal{G}_{\mathcal{V}}$ are different. Vertices with similar HOP vectors intuitively tend to be in the same cluster.

### 2.3 Weighted Projected Graph

Inspired by the foregoing intuition, we construct a weighted projected graph (WPG) that is a unipartite graph $\mathcal{G}_{\mathcal{V}}$ solely including all vertices in $\mathcal{V}$ via a projection over the input bipartite graph $\mathcal{G}$.

Specifically, we build the WPG $\mathcal{G}_{\mathcal{V}}$ with the edge weight matrix $\mathbf{W}_{\mathcal{V}}$ by a new edge-weighting scheme applied together with the one-mode projection [70]. To facilitate the design, we define the transition probabilities $p(u_i, v_j)$ from vertex $u_i$ to vertex $v_j$, and $p(v_j, u_i)$ from $v_j$ to $u_i$ on $\mathcal{G}$ as follows. Probability $p(u_i, v_j)$ (resp.

$p(v_j, u_i)$) connotes the one-hop jump probability of a random walk starting from $u_i$ to $v_j$ (resp. from $v_j$ to $u_i$).

$$p(u_i, v_j) = \frac{w(u_i, v_j)}{\sum_{v_l \in \mathcal{N}(u_i)} w(u_i, v_l)}, \quad p(v_j, u_i) = \frac{w(u_i, v_j)}{\sum_{u_h \in \mathcal{N}(v_j)} w(u_h, v_j)}. \quad (1)$$

For every two vertices $v_j, v_l \in \mathcal{V}$, we connect them in $\mathcal{G}_{\mathcal{V}}$ via an edge $(v_j, v_l)$ if both of them are connected to the same vertices in $\mathcal{U}$ of $\mathcal{G}$, *i.e.*, $\mathcal{N}(v_j) \cap \mathcal{N}(v_l) \neq \emptyset$. The edge weight $w_{\mathcal{V}}(v_j, v_l)$ is obtained as follows. On the input bipartite graph $\mathcal{G}$, apparently it requires a *two-hop jump* via a certain vertex $u_i \in \mathcal{N}(v_j) \cap \mathcal{N}(v_l)$ to jump from $v_j$ to $v_l$ with probability $p(v_j, u_i) \cdot p(u_i, v_l)$. Analogously, it needs a two-hop jump with probability $p(v_l, u_i) \cdot p(u_i, v_j)$ from $v_l$ to $v_j$ via $u_i$. Accordingly, the probability $q(v_j, u_i, v_l)$ that the two two-hop jumps both transit via the same vertex $u_i$ can be calculated by multiplying the two probabilities above, *i.e.*, $p(v_j, u_i) \cdot p(u_i, v_l)$. Since $v_j$ and $v_l$ may share multiple common neighbors $u_i \in \mathcal{N}(v_j) \cap \mathcal{N}(v_l)$ of $\mathcal{G}$, we aggregate all such probabilities $q(v_j, u_i, v_l)$ for $u_i \in \mathcal{U}$ together, resulting in the edge weight $w_{\mathcal{V}}(v_j, v_l)$ of edge $(v_j, v_l)$ in the WPG $\mathcal{G}_{\mathcal{V}}$, as shown in Eq. (2). Notice that a square root operator is applied for taking a geometric mean of $q(v_j, u_i, v_l)$ and $q(v_l, u_i, v_j)$ and meanwhile alleviating the issue of small values.

$$w_{\mathcal{V}}(v_j, v_l) = \sum_{u_i \in \mathcal{U}} \sqrt{p(v_j, u_i) \cdot p(u_i, v_l)} \cdot \sqrt{p(v_l, u_i) \cdot p(u_i, v_j)} \quad (2)$$

Notice that Eq. (2) can be reorganized as follows:

$$w_{\mathcal{V}}(v_j, v_l) = \sum_{u_i \in \mathcal{U}} \sqrt{p(v_j, u_i) \cdot p(u_i, v_j)} \cdot \sqrt{p(v_l, u_i) \cdot p(u_i, v_l)} \quad (3)$$

Let $\mathbf{Q} \in \mathbb{R}^{|\mathcal{V}| \times |\mathcal{U}|}$ be a matrix where the $(j, i)$-th element is $\mathbf{Q}_{j,i} = \sqrt{p(v_j, u_i) \cdot p(u_i, v_j)}$ if $(u_i, v_j) \in \mathcal{E}$, and 0 otherwise. Then Eq. (3) can be re-written as

$$w_{\mathcal{V}}(v_j, v_l) = \mathbf{Q}_j \cdot \mathbf{Q}_l, \quad (4)$$

and accordingly the edge weight matrix $\mathbf{W}_{\mathcal{V}}$ of $\mathcal{G}_{\mathcal{V}}$ is $\mathbf{W}_{\mathcal{V}} = \mathbf{Q}\mathbf{Q}^{\top}$.

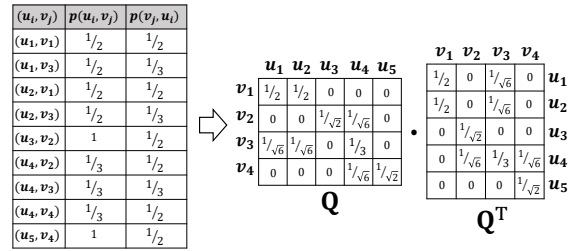

**Figure 3: Example $p(u_i, v_j)$ and $\mathbf{Q}$**

*Example 2.1.* Suppose that the edge weights $w(u_i, v_j)$ in the bipartite graph $\mathcal{G}$ of Figure 2 are all 1. Then, by Eq. (1), the one-hop jump probabilities $p(u_i, v_j)$ and $p(v_j, u_i)$ of every node pair in $\mathcal{U} \times \mathcal{V}$ are presented on the left of Figure 3. Then as an example, $\mathbf{Q}_{3,1} = \sqrt{p(v_3, u_1) \cdot (u_1, v_3)} = 1/\sqrt{6}$. After getting $\mathbf{Q}$ in Figure 3, the edge weights of $\mathcal{G}_{\mathcal{V}}$ is calculated by Eq. (4). For example, $w_{\mathcal{V}}(v_1, v_3) = \sum_{u_i \in \mathcal{U}} \mathbf{Q}_{1,i} \times \mathbf{Q}_{3,i} = \frac{1}{2} \times \frac{1}{\sqrt{6}} + \frac{1}{2} \times \frac{1}{\sqrt{6}} = \frac{1}{\sqrt{6}}$.

## 2.4 High-Order Perspective Vectors

In a bipartite graph $\mathcal{G}$, vertices $u_i$ and $u_h$ in $\mathcal{U}$ might have resembling or diverse neighbor sets $\mathcal{N}(u_i)$ and $\mathcal{N}(u_h)$, both of which are subsets of $\mathcal{V}$. If $\mathcal{N}(u_i)$ and $\mathcal{N}(u_h)$ are radically different, $u_i$ and $u_h$ yield disparate perspectives towards the WPG $\mathcal{G}_\mathcal{V}$. For instance, in Figure 2, vertex $u_1$ is adjacent to $v_1$ and $v_3$, while vertex $u_3$ connects to $v_2$, and consequently the HOP vectors of $u_1$ and $u_3$ are radically different. On the other hand, $u_1$ and $u_2$ share similar HOP vectors since they have similar perspectives towards WPG $\mathcal{G}_\mathcal{V}$.

Next, we present the formula to compute the HOP vectors for vertices $u_i \in \mathcal{U}$ towards the WPG $\mathcal{G}_\mathcal{V}$ with the consideration of high-order topological connections. For every vertex $u_i$, we first connect itself to all its neighbors $v_j \in \mathcal{N}(u_i)$ in the WPG $\mathcal{G}_\mathcal{V}$, with weight $p(u_i, v_j)$. Afterwards, we simulate *random walks with restart* [54] from every $u_i$ to the WPG $\mathcal{G}_\mathcal{V}$ to produce the HOP vector of $u_i$ w.r.t. $\mathcal{G}_\mathcal{V}$. We denote by $\mathbf{P} \in \mathbb{R}^{|\mathcal{U}| \times |\mathcal{V}|}$ the transition matrix comprising the one-hop transition probability for each edge $(u_i, v_j) \in \mathcal{E}$, $\mathbf{P}_{i,j} = p(u_i, v_j)$. Further, on the WPG $\mathcal{G}_\mathcal{V}$, $(\mathbf{W}_\mathcal{V})_{j,l}^\lambda$, *i.e.*, $(\mathbf{QQ}^\top)_{j,l}^\lambda$ quantifies the strength of $\lambda$-step connections between vertices $v_j$ and $v_l$. Thus, the total strength $\mathbf{F}_{i,j}$ of connecting $u_i$ and $v_j$ in $\mathcal{G}_\mathcal{V}$ via any vertex $v_l \in \mathcal{G}_\mathcal{V}$ through random walks is calculated by

$$\mathbf{F} = \sum_{\lambda=0}^{\infty} (1-\alpha)\alpha^\lambda \cdot \mathbf{P}(\mathbf{QQ}^\top)^\lambda, \tag{5}$$

where $\alpha$ is a random walk decay factor in $(0, 1)$.

Lemma 2.2 indicates that each entry in $\mathbf{F}$ is bounded by 1. Further, $\mathbf{F}$ converges to an exact solution, which will be theoretically analysed in Section 3 (see Lemma 3.1)[1].

**Lemma 2.2.** $\forall u_i \in \mathcal{U}, v_j \in \mathcal{V}, 0 \le \mathbf{F}_{i,j} \le \mathbf{P}_{i,j} \le 1$ *holds.*

$\mathbf{F}$ is a $|\mathcal{U}| \times |\mathcal{V}|$ matrix, wherein each row vector $\mathbf{F}_i$ accommodates the stopping probabilities of random walks starting from $u_i$ to all vertices $v$ in $\mathcal{G}_\mathcal{V}$. Distinct vertices $u_i$ and $u_j$ correspond to different row vectors $\mathbf{F}_i$ and $\mathbf{F}_j$. Finally, in Eq. (6), we get the HOP vector $\mathbf{H}_i$ of $u_i$, w.r.t. the WPG $\mathcal{G}_\mathcal{V}$, by applying $L_2$ normalization over $\mathbf{F}_i$ to ensure that $\mathbf{H}_i$ has a unit $L_2$ norm $\|\mathbf{H}_i\|_2 = 1$.

$$\mathbf{H}_i = \mathbf{F}_i / \|\mathbf{F}_i\|_2 \tag{6}$$

The HOP vector $\mathbf{H}_i$ summarizes high-order affinities between the given vertex $u_i$ and any vertex $v_j \in \mathcal{V}$ from the perspective of $u_i$ over the WPG $\mathcal{G}_\mathcal{V}$, and hence $\mathbf{H}_i$ can act as a structural representation of $u_i$.

## 2.5 Objective Function

As aforementioned, two vertices $u_i$ and $u_h$ with similar connections to the WPG $\mathcal{G}_\mathcal{V}$ are likely to have similar HOP vectors, and naturally, they tend to be in the same cluster, and *vice versa*. Building on this idea, we formulate our $k$-BGC objective using the HOP vectors of all vertices $u_i$ in $\mathcal{U}$ as follows. The objective in Eq. (7) is to identify $k$ disjoint clusters $\{C_1, C_2, \cdots, C_k\}$ such that the overall distance between HOP vectors of vertices in the same cluster and their mean is minimized, similar in spirit to K-Means [24].

$$\min_{\{C_1, \cdots, C_k\}} \sum_{j=1}^{k} \sum_{u_i \in C_j} \|\mathbf{H}_i - \overline{\mathbf{h}}^{(j)}\|_2^2, \tag{7}$$

---

[1]All proofs appear in Appendix A.

where $\overline{\mathbf{h}}^{(j)} = \frac{\sum_{u_i \in C_j} \mathbf{H}_i}{|C_j|}$ is the mean of HOP vectors of vertices in cluster $C_j$.

The objective poses two formidable challenges to address. First and foremost, the direct computation and materialization of the dense HOP matrix $\mathbf{H} \in \mathbb{R}^{|\mathcal{U}| \times |\mathcal{V}|}$ incur an exorbitant cost for large bipartite graphs with millions of nodes, given that $\mathbf{H}$ has a sophisticated definition that involves summing up an infinite series of matrices to capture high-order affinities as defined in Eq. (5) and (6). On top of that, the ultra-high dimensionality of $\mathbf{H}$ ($|\mathcal{V}|$ is often up to millions on large graphs) leads to numerous iterations till convergence in the optimization of Eq. (7).

To cope with these technical challenges, we first develop a base method HOPE in Section 3, which achieves superior effectiveness while being efficient. We further propose HOPE+ in Section 4 to circumvent certain limitations of HOPE and expedite the practical efficiency while gaining improved clustering quality.

## 3 THE HOPE METHOD

The workflow of HOPE for $k$-BGC is illustrated in Figure 4. More concretely, to mitigate the severe issue of materializing the HOP matrix $\mathbf{H}$ and direct clustering of $\mathbf{H}$, HOPE first derives a low-rank approximation $\mathbf{X} \in \mathbb{R}^{|\mathcal{U}| \times \beta}$ of $\mathbf{H} \in \mathbb{R}^{|\mathcal{U}| \times |\mathcal{V}|}$ ($\beta \ll |\mathcal{V}|$), such that for any two nodes $u_i$ and $u_j$ in $\mathcal{U}$, the difference between Euclidean distances $\|\mathbf{X}_i - \mathbf{X}_j\|_2$ and $\|\mathbf{H}_i - \mathbf{H}_j\|_2$ is bounded. Later, rather than $\mathbf{H}$ itself, $\mathbf{X}$ is used as a representation of $\mathbf{H}$ for clustering via optimizing the objective in Eq. (7). For instance, in Figure 4, our task to cluster five 4-dimensional HOP vectors of nodes $u_1$-$u_5$ turns to group five length-3 vectors into 2 clusters $C_1 = \{u_1, u_2\}$ and $C_2 = \{u_3, u_4, u_5\}$.

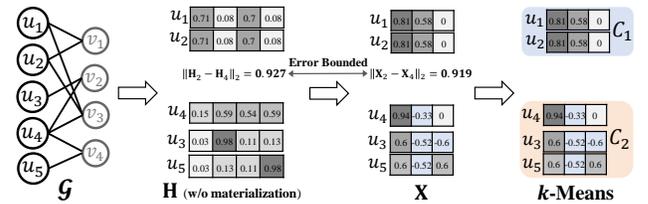

**Figure 4: A Running Example of HOPE**

However, computing such a low-rank representation $\mathbf{X}$ is highly challenging, especially on how to bound the difference between $\mathbf{X}$ and $\mathbf{H}$, without actually materializing $\mathbf{H}$ for the sake of efficiency. In the following, we first conduct theoretical analysis to derive $\mathbf{X}$, before elaborating on the algorithmic details and asymptotic performance of HOPE.

**Analysis to Compute X.** At first, we derive the following crucial property in Lemma 3.1 for $\sum_{\lambda=0}^{\infty} (1-\alpha)\alpha^\lambda (\mathbf{QQ}^\top)^\lambda$ in Eq. (5) by virtue of the symmetry of matrix $\mathbf{QQ}^\top$ and the semi-orthogonal property of singular vectors.

**Lemma 3.1.** *Let* $\mathbf{U}$ *and* $\Sigma$ *be the left singular vectors and singular values of* $\mathbf{Q}$. *Then,* $\mathbf{U}_{*,i}$ *and* $\frac{1-\alpha}{1-\alpha \cdot \Sigma_{i,i}^2}$ *are the $i$-largest eigenvector and eigenvalue of* $\sum_{\lambda=0}^{\infty} (1-\alpha)\alpha^\lambda (\mathbf{QQ}^\top)^\lambda$.

Lemma 3.1 indicates that the eigenvectors and eigenvalues of the matrix $\sum_{\lambda=0}^{\infty} (1-\alpha)\alpha^\lambda (\mathbf{QQ}^\top)^\lambda$ can be readily obtained if the singular vectors $\mathbf{U}$ and singular values $\Sigma$ of $\mathbf{Q}$ are given, meaning

---

**Algorithm 1:** HOPE

**Data:** Bipartite graph $\mathcal{G} = (\mathcal{U} \cup \mathcal{V}, \mathcal{E})$

**Parameters:** The decay factor $\alpha$, the number $k$ of clusters and dimensionality $\beta$

**Result:** $k$ cluster sets: $C_1, C_2, \cdots, C_k$

1 Perform $\beta$-truncated SVD over **Q**;

2 Let $\Sigma$ be the top-$\beta$ singular values of **Q** and **U** be the corresponding left singular vectors;

3 Calculate $\widehat{\mathbf{X}}$ according to Eq. (8);

4 Normalize $\widehat{\mathbf{X}}$ as **X** such that each row has a unit $L_2$ norm;

5 Invoke $k$-Means to cluster the rows of **X**;

6 Let $C_1, C_2, \cdots, C_k$ be the output of the $k$-Means;

7 **return** $C_1, C_2, \cdots, C_k$;

---

that **F** in Eq. (5), the vital ingredient of **H** in Eq. (6), can be calculated by $\mathbf{PU}\frac{1-\alpha}{1-\alpha\Sigma^2}\mathbf{U}^\top$, indicating the convergence of **F**. Let

$$\widehat{\mathbf{X}} = \mathbf{PU}\frac{1-\alpha}{1-\alpha\cdot\Sigma^2}. \tag{8}$$

Then, using the fact of $\mathbf{U}^\top\mathbf{U} = \mathbf{I}$, we can get

$$\widehat{\mathbf{X}}\widehat{\mathbf{X}}^\top = \mathbf{PU}\left(\frac{1-\alpha}{1-\alpha\Sigma^2}\right)^2\mathbf{U}^\top\mathbf{P}^\top = \mathbf{PU}\frac{1-\alpha}{1-\alpha\Sigma^2}\mathbf{U}^\top\mathbf{U}\frac{1-\alpha}{1-\alpha\Sigma^2}\mathbf{U}^\top\mathbf{P}^\top = \mathbf{FF}^\top,$$

as well as

$$\|\widehat{\mathbf{X}}_i\|_2 = \left(\widehat{\mathbf{X}}\widehat{\mathbf{X}}^\top\right)_{i,i} = \left((\mathbf{PU}\frac{1-\alpha}{1-\alpha\Sigma^2}\mathbf{U}^\top)\cdot(\mathbf{PU}\frac{1-\alpha}{1-\alpha\Sigma^2}\mathbf{U}^\top)^\top\right)_{i,i} = \|\mathbf{F}_i\|_2.$$

The above equations imply that **X** with each row $\mathbf{X}_i = \frac{\widehat{\mathbf{X}}_i}{\|\widehat{\mathbf{X}}_i\|_2}$ can be employed as the low-rank approximation of **H**, when **U** and $\Sigma$ are the top-$\beta$ ($\beta \ll |\mathcal{V}|$) singular vectors and values of **Q** respectively.

**Theorem 3.2.** *Given the above matrix* $\mathbf{X} \in \mathbb{R}^{|\mathcal{U}|\times\beta}$, $\forall u_i, u_j \in \mathcal{U}$, *we have* $2\left(1 - \frac{1}{\sqrt{1-\frac{\sigma}{\|\mathbf{F}_i\|_2}}\sqrt{1-\frac{\sigma}{\|\mathbf{F}_j\|_2}}}\right)\cdot\mathbf{H}_i\mathbf{H}_j - \frac{2\cdot\sigma}{\sqrt{\|\mathbf{F}_i\|_2^2-\sigma}\sqrt{\|\mathbf{F}_j\|_2^2-\sigma}} \leq \|\mathbf{X}_i - \mathbf{X}_j\|_2^2 - \|\mathbf{H}_i - \mathbf{H}_j\|_2^2 \leq 2\left(1 - \frac{1}{\sqrt{1+\frac{\sigma}{\|\mathbf{F}_i\|_2}}\sqrt{1+\frac{\sigma}{\|\mathbf{F}_j\|_2}}}\right)\cdot\mathbf{H}_i\mathbf{H}_j + \frac{2\cdot\sigma}{\sqrt{\|\mathbf{F}_i\|_2^2+\sigma}\sqrt{\|\mathbf{F}_j\|_2^2+\sigma}}$, *where* $\sigma = \left(\frac{1-\alpha}{1-\alpha\cdot\overline{\Sigma}_{\beta+1,\beta+1}^2}\right)^2$ *and* $\overline{\Sigma}_{\beta+1,\beta+1}$ *is the* $(\beta+1)$-*th largest singular value of* **Q**.

The accuracy guarantee between the low-rank approximation **X** and the HOP matrix **H** is established between $\|\mathbf{X}_i - \mathbf{X}_j\|_2^2$ and $\|\mathbf{H}_i - \mathbf{H}_j\|_2^2$ by Theorem 3.2. To demonstrate the empirical difference between $\|\mathbf{X}_i - \mathbf{X}_j\|_2^2$ and $\|\mathbf{H}_i - \mathbf{H}_j\|_2^2$, in Figure 5, we report the average relative error $\epsilon_r = \frac{1}{|\mathcal{U}|^2}\sum_{u_i, u_j \in \mathcal{U}}\frac{|\|\mathbf{X}_i - \mathbf{X}_j\|_2^2 - \|\mathbf{H}_i - \mathbf{H}_j\|_2^2|}{\|\mathbf{H}_i - \mathbf{H}_j\|_2^2}$ and absolute error $\epsilon_a = \frac{1}{|\mathcal{U}|^2}\sum_{u_i, u_j \in \mathcal{U}}|\|\mathbf{X}_i - \mathbf{X}_j\|_2^2 - \|\mathbf{H}_i - \mathbf{H}_j\|_2^2|$ when varying $\beta$. Note that $\epsilon_r \in [0, 1]$ and $\epsilon_a \in [0, 2]$. When increasing $\beta$ from 16 to 256, the errors $\epsilon_r$ and $\epsilon_a$ considerably diminish, and dwindle to small values around or below 0.1 when $\beta \geq 64$, which showcases that **X** can accurately approximate **H**. This observation matches the experimental results in Figure 10 where clustering quality improves and then becomes steady when increasing $\beta$.

**The HOPE Algorithm.** On the basis of the above analysis, we can obtain the low-rank approximation **X** of **H** through an efficient truncated *singular value decomposition* (SVD) over the sparse matrix **Q** using a number of fast matrix operations, without materializing **H** explicitly, thereby leading to a significant reduction in computation cost. Algorithm 1 displays the pseudo-code of HOPE, which begins with the input of a bipartite graph $\mathcal{G}$, decay factor $\alpha$, the number $k$ of clusters, and dimensionality $\beta$ that is usually a multiple of $k$. After that, HOPE performs a $\beta$-truncated SVD over matrix **Q** to get its top-$\beta$ left singular vectors **U** and a $\beta \times \beta$ diagonal matrix $\Sigma$ containing the top-$\beta$ singular values (Lines 1-2). At Line 3, we calculate the matrix $\widehat{\mathbf{X}}$ according to Eq. (8), and then normalize $\widehat{\mathbf{X}}$ as **X** such that each $i$-th row has a unit $L_2$ norm, *i.e.*, $\mathbf{X}_i = \frac{\widehat{\mathbf{X}}_i}{\|\widehat{\mathbf{X}}_i\|_2}$ (Line 4). Lastly, Algorithm 1 invokes $k$-Means algorithm [24] to cluster the rows of the low-rank approximation matrix **X** into clusters $C_1, C_2\cdots, C_k$ and return them as the final result (Lines 5-7).

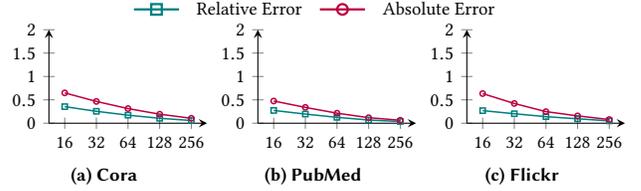

**Figure 5: Approximation errors of X when varying $\beta$.**

**Complexity.** Notice that both **P** and **Q** are sparse matrices with $|\mathcal{E}|$ non-zero entries. Hence, Line 1 in Algorithm 1 takes $O(|\mathcal{E}| \cdot \beta)$ time [49]. The computation of $\widehat{\mathbf{X}}$ by Eq. (8) requires a sparse matrix multiplication, incurring $O(|\mathcal{E}| \cdot \beta)$ time, and the normalization of $\widehat{\mathbf{X}}$ at Line 4 takes $O(|\mathcal{U}| \cdot \beta)$ time. As stated by [24], $k$-Means over **X** runs in $O(|\mathcal{U}| \cdot \beta \cdot k \cdot T)$ time, where $T$ stands for the number of iterations in $k$-Means. In total, the time complexity of HOPE is bounded by $O((|\mathcal{E}| + |\mathcal{U}| \cdot k) \cdot \beta)$. The space overhead of HOPE is chiefly determined by the sizes of $\widehat{\mathbf{X}}$ and **X**, containing $O(|\mathcal{U}| \cdot \beta)$ entries. Together with the matrices **P** and **Q**, the overall space cost entailed by HOPE is $O(|\mathcal{E}| + |\mathcal{U}| \cdot \beta)$.

## 4 THE HOPE+ METHOD

As revealed in experiments in Section 5, HOPE achieves superior clustering quality, while being efficient. Nevertheless, HOPE inherits the defects of $k$-Means, *e.g.*, the tendency to get stuck at local optima, and the numerous iterations required for convergence, especially on large bipartite graphs. To alleviate such issues, we further propose HOPE+ which is able to advance clustering performance in both efficiency and effectiveness.

In Section 4.1, we first explain how HOPE+ transforms and disassembles the $k$-BGC objective in Section 2.5 into two optimization problems solved one after another via efficient algorithmic designs, as elucidated in Section 4.2 and Section 4.3 respectively. Briefly, the $k$-BGC objective in Eq. (7) is converted to its equivalent form of a matrix trace maximization. The first problem is efficiently optimized with a fractional solution $\mathbf{L} \in \mathbb{R}^{|\mathcal{U}|\times k}$ derived by eigen-decomposition, when certain constraints are relaxed (Section 4.2). The second optimization problem is formulated to minimize the difference (quantified by the popular Frobenius norm and spectral norm) between a vertex-cluster membership indicator matrix (VCMI) **C** and **L**. Therefore, we devise two *rounding algorithms*,

FNEM and SNEM, to get **C** when minimizing errors in the aforementioned norms (Section 4.3). Section 4.4 includes the analysis of the overall time and space complexities of HOPE+ with FNEM and SNEM.

### 4.1 Two-Stage Optimizations

Recall that $\overline{\mathbf{h}}^{(j)} = \frac{\sum_{u_i \in C_j} \mathbf{H}_i}{|C_j|}$ in Eq. (7) and $\|\mathbf{H}_i\|_2 = 1$, $\forall u_i \in \mathcal{U}$ in Eq. (6), based on which, we establish the equivalence between the objective in Eq. (7) and the trace optimization in Eq. (9).

$$\min_{\{C_1,\cdots,C_k\}} \sum_{j=1}^{k} \sum_{u_i \in C_j} \|\mathbf{H}_i - \overline{\mathbf{h}}^{(j)}\|_2^2$$

$$= \min_{\{C_1,\cdots,C_k\}} \sum_{j=1}^{k} \sum_{u_i \in C_j} \|\mathbf{H}_i\|_2^2 - \sum_{j=1}^{k} \sum_{u_i, u_l \in C_j} \frac{\mathbf{H}_i \cdot \mathbf{H}_l}{|C_j|}$$

$$\iff \max_{\mathbf{C}} \mathrm{Tr}(\mathbf{C}^\top \mathbf{H}\mathbf{H}^\top \mathbf{C}) \tag{9}$$

where $\mathbf{C} \in \mathbb{R}^{|\mathcal{U}| \times k}$ in Eq. (9) is the vertex-cluster membership indicator matrix (VCMI) with each $(i, j)$-th entry to be

$$\mathbf{C}_{i,j} = \begin{cases} \frac{1}{\sqrt{|C_j|}} & \text{if } u_i \in C_j, \\ 0 & \text{otherwise,} \end{cases} \tag{10}$$

where $|C_j|$ is the size of cluster $C_j$.

In Eq. (10), observe that **C** is required to fulfill two requirements: (i) each row in **C** contains only one non-zero and positive entry; and (ii) each column has a unit $L_2$ norm in which all entries are equal. Together these two constraints ensure that **C** has orthogonal columns, *i.e.*, $\mathbf{C}^\top \mathbf{C} = \mathbf{I}$. Using Ky Fan's trace maximization principle in Lemma 4.1, the optimal solution to Eq. (9) is the $k$-largest eigenvectors of $\mathbf{H}\mathbf{H}^\top$ when the two foregoing constraints on **C** are relaxed to satisfying $\mathbf{C}^\top \mathbf{C} = \mathbf{I}$.

**Lemma 4.1 (Ky Fan's trace maximization principle [15]).** *Given a symmetric real matrix* $\mathbf{M} \in \mathbb{R}^{n \times n}$ *with distinct eigenvalues* $\psi_1(\mathbf{M})$, $\psi_2(\mathbf{M})$, $\cdots$, $\psi_n(\mathbf{M})$, *sorted by algebraic value in descending order, eigenvectors* $\mathbf{L}$ *and integer* $k \le n$, *we have*

$$\max_{\mathbf{Y}^\top \mathbf{Y} = \mathbf{I}_k} \mathrm{Tr}(\mathbf{Y}^\top \mathbf{M}\mathbf{Y}) = \mathrm{Tr}(\mathbf{L}^\top \mathbf{M}\mathbf{L}) = \sum_{i=1}^{k} \psi_i(\mathbf{M}).$$

According to Lemma 4.1, we can utilize the $k$-largest eigenvectors **L** of $\mathbf{H}\mathbf{H}^\top$ as a solution in $\mathbb{R}^{|\mathcal{U}| \times k}$ to optimize the trace maximization problem in Eq. (9). Afterwards, a VCMI **C** can be obtained by minimizing the differences between **C** and **L** via the optimization problem formulated in Eq. (11).

$$\min_{\mathbf{C}} |\mathrm{Tr}(\mathbf{L}^\top \mathbf{H}\mathbf{H}^\top \mathbf{L}) - \mathrm{Tr}(\mathbf{C}^\top \mathbf{H}\mathbf{H}^\top \mathbf{C})| \tag{11}$$
$$\text{s.t. Eq. (10) holds } \forall u_i \in \mathcal{U} \text{ and } 1 \le j \le k$$

Summing up, we transform the $k$-BGC objective in Eq. (7) into two optimization problems in Eq. (9) and (11), to be solved in order.

**The HOPE+ Algorithm.** The pseudo-code of HOPE+ is outlined in Algorithm 2, which proceeds in two stages. The first stage is to optimize Eq. (9) via approximate $k$-largest eigenvectors **L** by Lemma 4.1 (Lines 1-5). Note that the construction of matrix $\mathbf{H}\mathbf{H}^\top$ entails a quadratic space overhead of $O(|\mathcal{U}|^2)$ and $O(|\mathcal{U}|^2 \cdot |\mathcal{V}| + |\mathcal{E}| \cdot |\mathcal{V}|)$ computation time, which is intolerable for massive bipartite graphs. To mitigate this issue, in Section 4.2, we efficiently approximate $k$-largest Eigenvectors **L** without explicitly materializing $\mathbf{H}\mathbf{H}^\top$, and we will explain Lines 1-5 shortly. With **L** at hand, HOPE+ then enters into the second stage, which attends to generating the final clustering result, namely VCMI **C**, based on the optimization problem in Eq. (11). Prior to that, **C** is initialized via a simple greedy seeding strategy (Lines 6-10). To be specific, for every vertex $u_i \in \mathcal{U}$, we first get the column index $j^*$ corresponding to the maximum entry $\mathbf{L}_{i,j^*}$ in $\mathbf{L}_i$ (Line 8), and assign $u_i$ into cluster $C_j$ (Line 9). This greedy seeding strategy provides us a high-quality initialization of **C** to facilitate the convergence of **C**, when dealing with the optimization problem in Eq. (11). However, it is highly challenging to solve Eq. (11) as it may involve prohibitively expensive time of up to $O(|\mathcal{U}|^2 k)$. In lieu of directly working on Eq. (11), in Section 4.3, we conduct further analysis to transform Eq. (11) to its equivalent but simplified form, and present two rounding algorithms for the refinement and obtainment of **C** (Line 11), namely FNEM and SNEM under the settings of the popular Frobenius norm error and spectral norm error, respectively. At Line 12, **C** is returned as the final clustering result.

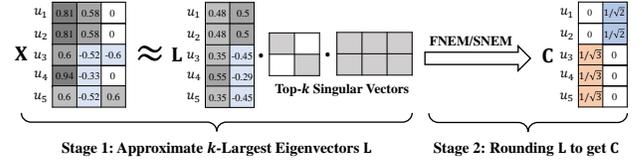

**Figure 6: A Running Example of HOPE+**

*Example 4.2.* Figure 6 presents a running example of HOPE+ over the bipartite graph $\mathcal{G}$ with $k = 2$ in Figure 4 for the clustering of five nodes $u_1$-$u_5$. Without materializing **H** and computing $\mathbf{H}\mathbf{H}^\top$, HOPE+ first transforms the problem of calculating the 2-largest eigenvectors of $\mathbf{H}\mathbf{H}^\top$ as computing the top-2 singular vectors **L** of **X**, the 3-dimension approximation of **H** obtained as in Algorithm 1. After that, HOPE+ conducts a rounding procedure in Algorithm 3 over **L** $\in \mathbb{R}^{5 \times 2}$ to get a VCMI matrix **C**, in which each row contains only one non-zero entry indicating the cluster membership of the node corresponded by the row. For example, in Figure 6, rows 1 and 2 in **C** have non-zero elements $\frac{1}{\sqrt{2}}$ at the second column, while the entries at the first columns of rows 3-4 are nonzero, i.e., $\frac{1}{\sqrt{3}}$, resulting in two clusters $\{u_1, u_2\}$ and $\{u_3, u_4, u_5\}$. In HOPE+, rounding the top-2 singular vectors **L** to get the VCMI **C** for producing clusters averts the high computationl costs entailed by using the $k$-Means in HOPE.

### 4.2 Approximating $k$-Largest Eigenvectors **L**

As alluded to earlier, in the first stage of HOPE+, the goal is to efficiently approximate the top-$k$ eigenvectors **L** to solve Eq. (9) based on Lemma 4.1, but without materializing $\mathbf{H}\mathbf{H}^\top$.

Recall that in Lemma 3.1, we have $\mathbf{X}\mathbf{X}^\top = \mathbf{H}\mathbf{H}^\top$ when $\beta = |\mathcal{V}|$, whereby we can obtain a low-rank approximation $\mathbf{X} \in \mathbb{R}^{|\mathcal{U}| \times \beta}$ ($k < \beta \ll |\mathcal{V}|$) of **H** efficiently such that $\mathbf{X}\mathbf{X}^\top \approx \mathbf{H}\mathbf{H}^\top$. Further,

---

**Algorithm 2:** HOPE+

---

**Data:** Bipartite graph $\mathcal{G} = (\mathcal{U} \cup \mathcal{V}, \mathcal{E})$
**Parameters:** The decay factor $\alpha$, the number $k$ of clusters, dimensionality $\beta$, and the number $T$ of iterations

**Result:** The final VCMI C

/* Stage 1: Approximate $k$-largest eigenvectors L */

Lines 1-4 are the same as Lines 1-4 in Algorithm 1;

5 Perform the $k$-truncated SVD over X to get the top-$k$ left singular vectors L to solve Eq. (9);

/* Stage 2: Rounding for C */

6 Initialize an empty VCMI $C \leftarrow \mathbf{0}^{|\mathcal{U}| \times k}$;

7 **for** $u_i \in \mathcal{U}$ **do**

8 $\quad j^* = \arg\max_{1 \le j \le k} L_{i,j}$;

9 $\quad C_{i,j^*} \leftarrow 1$;

10 Normalize C such that each column has a unit $L_2$ norm;

11 Invoke Algorithm 3 for FNEM/SNEM rounding between L and C for at most $T$ iterations;

12 **return** C;

---

by leveraging Lemma 4.3, calculating the $k$-largest eigenvectors of $HH^\top$ is reduced to obtaining the top-$k$ left singular vectors of the low-rank approximation X.

**Lemma 4.3** ([49]). *Given an SVD of matrix* M, *the top-$k$ left singular vectors of* M *are the $k$ largest eigenvectors of* $MM^\top$.

This idea is realized at Lines 1-5 in Algorithm 2 with a detailed procedure of approximating the $k$-largest eigenvectors of $HH^\top$. More precisely, Algorithm 2 initially generates the $|\mathcal{U}| \times \beta$-sized low-rank X of the HOP matrix H, identical to Lines 1-4 in Algorithm 1. Subsequently, at Line 5 in Algorithm 2, a $k$-truncated SVD is conducted over X to produce the top-$k$ left singular vectors of X, which are returned as the approximate $k$-largest eigenvectors L of $HH^\top$, according to Lemma 4.2 and the fact $XX^\top \approx HH^\top$.

Notice that Lines 1-4 in Algorithm 2 run in $O((|\mathcal{U}| + |\mathcal{E}|) \cdot \beta)$ time, and the $k$-truncated SVD at Line 5 in Algorithm 2 is performed over a $|\mathcal{U}| \times \beta$ matrix, leading to $O(|\mathcal{U}| \cdot \beta^2)$ time. X and other intermediate matrices in Algorithm 2 consume $O(|\mathcal{U}| \cdot \beta)$ space. In sum, Lines 1-5 in Algorithm 2 avoid materializing $HH^\top$ and enable a fast computation of L with $O(|\mathcal{U}| \cdot \beta^2 + |\mathcal{E}| \cdot \beta)$ time and $O(|\mathcal{E}| + |\mathcal{U}| \cdot \beta)$ space, both of which are linear to graph size.

### 4.3 Rounding Algorithms

As explained, after obtaining the approximate L above, the second stage of HOPE+ is to derive C by solving Eq. (11). For efficiency purpose, instead of directly working on Eq. (11), we convert it to its equivalent form that is tractable to solve via the following analysis.

In particular, we first capitalize on the *cyclic property* of matrix trace to get $\text{Tr}(C^\top HH^\top C) = \text{Tr}(HH^\top CC^\top)$ and $\text{Tr}(L^\top HH^\top L) = \text{Tr}(HH^\top LL^\top)$, so as to transform Eq. (11) into

$$\min_{C} \text{Tr}(|HH^\top (LL^\top - CC^\top)|) \tag{12}$$

s.t. Eq. (10) holds $\forall u_i \in \mathcal{U}$ and $1 \le j \le k$.

---

**Algorithm 3:** FNEM/SNEM Rounding

---

**Data:** The initial VCMI C and $k$-largest eigenvectors L
**Parameters:** The number $T$ of iterations and *URT*
**Result:** The final VCMI C

1 $t \leftarrow 1$;

2 **while** C *does not converges and* $t \le T$ **do**

3 $\quad$ /* Update T when fixing C */

$\quad$ **switch** *URT* **do**

4 $\quad\quad$ **case** FNEM **do**

5 $\quad\quad\quad$ Perform a full SVD over $L^\top C$ to get the left and right singular vectors $\Phi$ and $\Psi$;

6 $\quad\quad\quad T \leftarrow \Phi\Psi^\top$;

7 $\quad\quad$ **case** SNEM **do** $T \leftarrow L^\top C$;

$\quad$ /* Update C when fixing T */

8 $\quad C \leftarrow \mathbf{0}$;

9 $\quad$ **for** $u_i \in \mathcal{U}$ **do**

10 $\quad\quad j^* = \arg\max_{1 \le j \le k} (LT)_{i,j}$;

11 $\quad\quad C_{i,j^*} \leftarrow 1$;

12 $\quad$ Normalize C such that each column has a unit $L_2$ norm;

13 $\quad t \leftarrow t + 1$;

14 **return** C;

---

If there exists a matrix $T \in \mathbb{R}^{k \times k}$ that minimizes certain matrix norm error (*e.g.*, Frobenius norm or spectral norm) between LT and C as in Eq. (13), then Eq. (12) is optimized since $CC^\top \approx LT \cdot (LT)^\top = LTT^\top L^\top = LL^\top$.

$$\min_{T,C} \|LT - C\|_* \ s.t. \ TT^\top = I \tag{13}$$

As such, the optimization problem in Eq. (11) turns into Eq. (13), which can be efficiently solved as shown shortly. The Frobenius norm $\| \cdot \|_F$ and spectral norm $\| \cdot \|_2$ are two popular choices for the matrix norm $\| \cdot \|_*$ in Eq. (13). In what follows, we develop two fast rounding algorithms, namely *FNEM[2] Rounding* and *SNEM[3] Rounding*, to efficiently solve Eq. (13) to get C, when either Frobenius norm or spectral norm is applied. Algorithm 3 illustrates the pseudo-code of FNEM or SNEM rounding, whose input parameters include an initial VCMI C, the $k$-largest eigenvectors L, the total number of iterations $T$, and a parameter *URT* (short for Updating Rule for T) to specify which rounding scheme is adopted, *i.e.*, FNEM or SNEM.

**FNEM Rounding.** We first delineate Algorithm 3 under the FNEM rounding scheme. From Lines 1 to 13, FNEM rounding works in an iterative process to refine T and C in an alternative fashion until C converges (i.e., C remains unchanged) or $T$ iterations are completed. In each iteration, we fix one of T and C, and update the other.

Now we explain the updating rules in detail. Note that the objective function in Eq. (13) can be regarded as an *orthogonal Procrustes problem* [22] when C is fixed. To be specific, given two matrices L and C, the orthogonal Procrustes problem asks to find an orthogonal matrix T ($i.e.$, $TT^\top = I$) which most closely maps L to C such that $\|LT - C\|_F$ is minimized. By leveraging our result in Lemma

---

4.4, the SVD result of $\mathbf{L}^\top \mathbf{C}$ can be utilized to infer the optimal $\mathbf{T}$, *i.e.*, $\mathbf{T} = \mathbf{\Phi}\mathbf{\Psi}^\top$ as Lines 5-6 in Algorithm 3. Since $\mathbf{L}^\top \mathbf{C}$ is a $k \times k$ matrix and $k$ is usually small, the SVD can be done efficiently.

LEMMA 4.4. *Let $\mathbf{\Phi}$ and $\mathbf{\Psi}$ be the left and right singular vectors of $\mathbf{L}^\top \mathbf{C}$, respectively. Then $\mathbf{T}^* = \arg\min_{\mathbf{T}} \|\mathbf{LT} - \mathbf{C}\|_F^2 = \mathbf{\Phi}\mathbf{\Psi}^\top$.*

After obtaining $\mathbf{T}$ when $\mathbf{C}$ is fixed as described above, we then need to fix $\mathbf{T}$ and update $\mathbf{C}$. The problem in Eq. (13) turns into

$$\min_{\mathbf{C}} \|\mathbf{LT} - \mathbf{C}\|_F^2. \tag{14}$$

Recall that $\mathbf{C}$ is a VCMI matrix defined in Eq. (10), wherein every $i$-th row has only one non-zero entry. Simply, to optimize Eq. (14), we can determine the cluster id $C_{j^*}$ for each vertex $u_i \in \mathcal{U}$ (Lines 8-11 in Algorithm 3), by locating the column $j^*$ with the maximum entry $(\mathbf{LT})_{i,j^*}$ in the $i$-th row of $\mathbf{LT}$ (Line 10) and then setting $\mathbf{C}_{i,j^*}$ to 1 (Line 8). After processing all vertices in $\mathcal{U}$, each column of $\mathbf{C}$ is $L_2$-normalized at Line 12 so as to ensure Eq. (10). At last, FNEM rounding increases $t$ by 1 at Line 13 and executes the next iteration.

**SNEM Rounding.** When spectral norm is considered in Eq. (13) (i.e., $URT$ is set to SNEM) rounding, the optimization problem becomes $\min_{\mathbf{T},\mathbf{C}} \|\mathbf{LT} - \mathbf{C}\|_2$ *s.t.* $\mathbf{TT}^\top = \mathbf{I}$. We also adopt the same iterative process for SNEM as in Algorithm 3, where we update $\mathbf{T}$ and $\mathbf{C}$ alternatively in an iteration. When $\mathbf{C}$ is fixed, SNEM rounding is to find matrix $\mathbf{T}$ that solves

$$\min_{\mathbf{T}} \|\mathbf{LT} - \mathbf{C}\|_2. \tag{15}$$

The optimal solution to Eq. (15) is simply $\mathbf{T}^* = \mathbf{L}^\top \mathbf{C}$ as per our carefully-analyzed result in Lemma 4.5.

LEMMA 4.5. $\mathbf{T}^* = \arg\min_{\mathbf{T}} \|\mathbf{LT} - \mathbf{C}\|_2 = \mathbf{L}^\top \mathbf{C}$.

In Algorithm 3, in every iteration, when $\mathbf{C}$ is fixed, the updating rule of $\mathbf{T}$ is to set it as $\mathbf{L}^\top \mathbf{C}$ for SNEM rounding (Line 7), and meanwhile, the rule for updating $\mathbf{C}$ with a given $\mathbf{T}$ in SNEM rounding is consistent with that in FNEM rounding. In comparison with FNEM rounding, SNEM rounding solely requires a simple matrix multiplication $\mathbf{L}^\top \mathbf{C}$ with $O(|\mathcal{U}| \cdot k)$ time for updating $\mathbf{T}$ (Line 7 in Algorithm 3), whereas FNEM rounding needs to conduct a full SVD over $\mathbf{L}^\top \mathbf{C}$ (Line 5) that costs $O(|\mathcal{U}| \cdot k^2 \cdot T_S)$ time, where $T_S$ is number of iterations needed in SVD and its typical setting could be as large as a few dozens in practice.

### 4.4 Complexity Analysis

Here we analyze the overall complexities of HOPE+. We dub HOPE+ with FNEM and SNEM rounding as HOPE+ (FNEM) and HOPE+ (SNEM), respectively. For FNEM rounding, it takes $O(|\mathcal{U}| \cdot k^2)$ time for the SVD and matrix multiplication at Lines 5-6 of each iteration (Algorithm 3). For SNEM rounding, in each iteration, the matrix $\mathbf{T}$ is updated at Line 7 in Algorithm 3, which can be done in $O(|\mathcal{U}| \cdot k)$ time since $\mathbf{C}$ is a sparse matrix with $|\mathcal{U}|$ non-zero entries. Both HOPE+ (FNEM) and HOPE+ (SNEM) share the same updating rule of $\mathbf{C}$ at Lines 8-12 per iteration, with $O(|\mathcal{U}| \cdot k)$ time to inspect the maximum element per row of a $|\mathcal{U}| \times k$ matrix. Further, recall that obtaining the top-$k$ eigenvectors of $\mathbf{HH}^\top$ in Section 4.2 takes $O(|\mathcal{U}| \cdot \beta^2 + |\mathcal{E}| \cdot \beta)$ time and $O(|\mathcal{E}| + |\mathcal{U}| \cdot \beta)$ space. Therefore, HOPE+ (FNEM) runs in $O(|\mathcal{E}| \cdot \beta + |\mathcal{U}| \cdot \beta^2 + |\mathcal{U}| \cdot k^2 T)$ time and consumes $O(|\mathcal{E}| + |\mathcal{U}| \cdot \beta)$ space; the time complexity and space complexity

### Table 2: Statistics of bipartite graphs. ($K = 10^3$, $M = 10^6$, $B = 10^9$)

| Dataset | $|\mathcal{U}|$ | $|\mathcal{V}|$ | $|\mathcal{E}|$ | Type | #clusters in $\mathcal{U}$ |
|---|---|---|---|---|---|
| CORA | 2.7K | 1.4K | 49.2K | unweighted | 7 |
| CiteSeer | 3.3K | 3.7K | 105.2K | unweighted | 6 |
| Flickr | 7.6K | 12K | 182.5K | unweighted | 9 |
| BlogCatalog | 5.2K | 8.2K | 369.4K | unweighted | 6 |
| PubMed | 19.7K | 0.5K | 988K | weighted | 3 |
| CORA-F | 19.8K | 8.7K | 1.13M | unweighted | 70 |
| LastFM (Asia) | 7.6K | 7.8K | 3.01M | unweighted | 18 |
| MIND | 94.4K | 711.2K | 16.5M | weighted | 18 |
| LastFM | 359.4K | 160.2K | 17.6M | weighted | 239 |
| MAG | 10.5M | 2.78M | 1.1B | weighted | 8 |

### Table 3: Evaluated methods.

| Algorithm | Category | Time complexity |
|---|---|---|
| LEADINGEIGENVECTOR (LE) [39] | Graph Clustering | $O((|\mathcal{U}| + |\mathcal{V}|)^2 + |\mathcal{E}|)$ |
| GIRVAN–NEWMAN [20] | | $O(|\mathcal{U}| \cdot |\mathcal{E}|^2)$ |
| SPECTRALCLUSTERING (SC) [55] | | $O(k \cdot |\mathcal{U}|^2)$ |
| NRP [64] | | $O(k \cdot (|\mathcal{E}| + k \cdot |\mathcal{U}|) \cdot \log(|\mathcal{U}|))$ |
| PPR [56] | | $O(|\mathcal{E}| \cdot (|\mathcal{U}| + |\mathcal{V}|) + k \cdot |\mathcal{U}| \cdot |\mathcal{V}|)$ |
| K-MEANS [24] | Data Clustering | $O(k \cdot |\mathcal{U}| \cdot |\mathcal{V}|)$ |
| K-MEDOIDS [29] | | $O(k \cdot |\mathcal{U}|^2 \cdot |\mathcal{V}|)$ |
| BIRCH [69] | | $O(|\mathcal{V}| \cdot |\mathcal{U}| \log(|\mathcal{U}|))$ |
| NMF [61] | | $O((|\mathcal{E}| + |\mathcal{U}| + |\mathcal{V}|) \cdot k)$ |
| SBC [31] | BGC | $O((|\mathcal{E}| + |\mathcal{U}| \cdot k + |\mathcal{V}| \cdot k) \cdot k)$ |
| SCC [12] | | $O((|\mathcal{E}| + |\mathcal{U}| \cdot k + |\mathcal{V}| \cdot k) \cdot \log k)$ |
| BISBM-KL [32] | | $O((|\mathcal{U}| + |\mathcal{V}|) \cdot k^2)$ |
| BISBM-MCMC [67] | | $O((|\mathcal{U}| + |\mathcal{V}|) \cdot k + |\mathcal{E}| \cdot \log^2(|\mathcal{U}| + |\mathcal{V}|))$ |
| HOPE | Our Solutions | $O((|\mathcal{E}| + |\mathcal{U}| \cdot k) \cdot \beta)$ |
| HOPE+ (FNEM) | | $O(|\mathcal{E}| \cdot \beta + |\mathcal{U}| \cdot \beta^2 + |\mathcal{U}| \cdot k^2)$ |
| HOPE+ (SNEM) | | $O(|\mathcal{E}| \cdot \beta + |\mathcal{U}| \cdot \beta^2 + |\mathcal{U}| \cdot k)$ |

of HOPE+ (SNEM) are bounded by $O(|\mathcal{E}| \cdot \beta + |\mathcal{U}| \cdot \beta^2 + |\mathcal{U}| \cdot kT)$ and $O(|\mathcal{E}| + |\mathcal{U}| \cdot \beta)$, respectively.

## 5 EXPERIMENTS

We experimentally evaluate HOPE and HOPE+ with FNEM and SNEM against 13 competitors on 10 real-world bipartite graph datasets, in terms of clustering quality and efficiency. The experiments run atop a Linux machine with 2 Intel(R) Xeon(R) Gold 6330 2.00GHz CPUs with 28 cores and 2 TB RAM. The reported results are averaged over 5 runs. For reproducibility, we publish all the codes and datasets at https://github.com/HKBU-LAGAS/HOPE.

### 5.1 Experimental Setup

**Datasets.** Table 2 lists the statistics of the 10 datasets in experiments. There are 5 small or medium-sized graphs and 5 large graphs with millions of edges, up to 1.1 billion edges in MAG. Both CORA and CiteSeer datasets [19] contain scientific publications $\mathcal{U}$ with edges to the corresponding keywords $\mathcal{V}$ in the publication, and the publications are in 7 and 6 classes (fields of study), respectively. CORA-F [5] is the full version of the CORA network. In Flickr [26], vertices in $\mathcal{U}$ (resp. $\mathcal{V}$) represent users (resp. tags) and edges specify the interest of users in tags, and users are in groups as clusters. In BlogCatalog [53], $\mathcal{U}$ is a set of bloggers, and $\mathcal{V}$ are the keywords generated from the bloggers' blogs. The labels of vertices in $\mathcal{U}$ signify the topic categories provided by the bloggers. The PubMed [47] dataset consists of scientific publications from the PubMed database pertaining to diabetes, classified into three categories, and each edge weight is the TF/IDF weight of the word in the publication. The MIND [58] dataset contains news articles $\mathcal{U}$ and users $\mathcal{V}$, and each edge is a click of a user on a news article, and the clustering labels are the categories of the news [63]. LastFM (Asia) [45] and LastFM [9] are from the LastFM music website. $\mathcal{U}$ (resp. $\mathcal{V}$) in LastFM (Asia) includes users from Asian countries (resp. artists liked by the users), whereas LastFM contains the play count of each music by each user. The clustering labels in both datasets are the locations of users. The MAG dataset is extracted from the Microsoft Academic Graph [48] by [6], with vertices in $\mathcal{U}$ and $\mathcal{V}$ representing papers and words in the abstracts, respectively, and each edge weight reflects the word occurrence of a word in a paper [63]. The labels correspond to 8 fine-grained fields of study.

**Table 4: Clustering quality on small or medium-size datasets.**

| Algorithm | CORA | | | | CiteSeer | | | | BlogCatalog | | | | Flickr | | | | PubMed | | | | Rank |
|---|---|---|---|---|---|---|---|---|---|---|---|---|---|---|---|---|---|---|---|---|---|
| | Acc | F1 | NMI | ARI | Acc | F1 | NMI | ARI | Acc | F1 | NMI | ARI | Acc | F1 | NMI | ARI | Acc | F1 | NMI | ARI | |
| LE | 0.362 | 0.182 | 0.141 | 0.138 | 0.311 | 0.181 | 0.086 | 0.078 | 0.351 | 0.242 | 0.176 | 0.112 | - | - | - | - | 0.591 | 0.585 | 0.249 | 0.233 | 11.5 |
| Girvan–Newman | 0.302 | 0.068 | 0.003 | 0 | 0.212 | 0.061 | 0 | 0 | - | - | - | - | - | - | - | - | - | - | - | - | 15.35 |
| SC | 0.299 | 0.082 | 0.015 | 0.001 | 0.215 | 0.073 | 0.010 | 0 | 0.203 | 0.120 | 0.032 | 0.006 | 0.125 | 0.041 | 0.015 | 0 | 0.606 | 0.601 | 0.312 | 0.28 | 12.05 |
| K-Means | 0.342 | 0.300 | 0.197 | 0.098 | 0.435 | 0.410 | 0.206 | 0.159 | 0.263 | 0.233 | 0.155 | 0.048 | 0.133 | 0.054 | 0.03 | 0 | 0.595 | 0.581 | 0.311 | 0.28 | 9.5 |
| K-Medoids | 0.345 | 0.269 | 0.123 | 0.088 | 0.312 | 0.266 | 0.050 | 0.047 | 0.258 | 0.244 | 0.069 | 0.041 | 0.118 | 0.025 | 0 | 0 | 0.396 | 0.381 | 0.012 | 0.015 | 13.15 |
| NMF | 0.376 | 0.340 | 0.162 | 0.113 | 0.499 | 0.451 | 0.224 | 0.226 | 0.368 | 0.313 | 0.219 | 0.128 | 0.267 | 0.211 | 0.163 | 0.115 | 0.6 | 0.593 | 0.307 | 0.279 | 7.55 |
| Birch | 0.371 | 0.327 | 0.213 | 0.097 | 0.468 | 0.453 | 0.226 | 0.17 | 0.395 | 0.295 | 0.295 | 0.147 | 0.121 | 0.033 | 0.013 | 0 | 0.399 | 0.190 | 0 | 0 | 10.25 |
| NRP | 0.387 | 0.333 | 0.205 | 0.123 | 0.43 | 0.421 | 0.182 | 0.145 | 0.486 | 0.491 | 0.278 | 0.212 | 0.411 | 0.373 | 0.285 | 0.205 | 0.604 | 0.614 | 0.233 | 0.221 | 7.5 |
| PPR | 0.44 | 0.404 | 0.251 | 0.176 | 0.627 | 0.573 | 0.36 | 0.357 | 0.451 | 0.443 | 0.331 | 0.24 | 0.37 | 0.391 | 0.373 | 0.269 | 0.601 | 0.59 | 0.306 | 0.269 | 5.6 |
| SBC | 0.327 | 0.258 | 0.140 | 0.082 | 0.397 | 0.301 | 0.163 | 0.136 | 0.262 | 0.137 | 0.079 | 0.037 | 0.196 | 0.115 | 0.12 | 0.031 | 0.569 | 0.552 | 0.257 | 0.227 | 11.45 |
| SCC | 0.333 | 0.233 | 0.150 | 0.047 | 0.372 | 0.296 | 0.180 | 0.087 | 0.342 | 0.229 | 0.360 | 0.131 | 0.231 | 0.144 | 0.176 | 0.054 | 0.612 | 0.612 | 0.248 | 0.218 | 10.1 |
| biSBM-KL | 0.414 | 0.383 | 0.239 | 0.175 | 0.424 | 0.4 | 0.244 | 0.21 | 0.569 | 0.557 | 0.449 | 0.381 | 0.632 | 0.62 | 0.518 | 0.459 | 0.447 | 0.447 | 0.112 | 0.084 | 7.45 |
| biSBM-MCMC | 0.48 | 0.392 | 0.277 | 0.218 | 0.506 | 0.446 | 0.283 | 0.232 | 0.674 | 0.665 | 0.528 | 0.475 | 0.758 | 0.743 | 0.662 | 0.613 | 0.465 | 0.449 | 0.119 | 0.097 | 5.65 |
| HOPE | 0.604 | 0.55 | 0.376 | 0.323 | 0.663 | 0.618 | 0.397 | 0.395 | 0.698 | 0.667 | 0.543 | 0.5 | 0.742 | 0.739 | 0.638 | 0.583 | 0.626 | 0.634 | 0.267 | 0.246 | 2.5 |
| HOPE+ (FNEM) | 0.566 | 0.521 | 0.33 | 0.282 | 0.647 | 0.612 | 0.377 | 0.386 | 0.803 | 0.8 | 0.61 | 0.603 | 0.786 | 0.779 | 0.613 | 0.607 | 0.621 | 0.629 | 0.255 | 0.236 | 2.95 |
| HOPE+ (SNEM) | 0.607 | 0.526 | 0.356 | 0.319 | 0.682 | 0.588 | 0.392 | 0.411 | 0.808 | 0.801 | 0.61 | 0.611 | 0.793 | 0.748 | 0.675 | 0.658 | 0.623 | 0.627 | 0.262 | 0.236 | 2 |

**Table 5: Clustering quality on large datasets.**

| Algorithm | CORA-F | | | | LastFM (Asia) | | | | MIND | | | | LastFM | | | | MAG | | | | Rank |
|---|---|---|---|---|---|---|---|---|---|---|---|---|---|---|---|---|---|---|---|---|---|
| | Acc | F1 | NMI | ARI | Acc | F1 | NMI | ARI | Acc | F1 | NMI | ARI | Acc | F1 | NMI | ARI | Acc | F1 | NMI | ARI | |
| LE | 0.06 | 0.09 | 0.034 | 0.005 | - | - | - | - | - | - | - | - | - | - | - | - | - | - | - | - | 9 |
| SC | 0.256 | 0.201 | 0.319 | 0.127 | 0.16 | 0.053 | 0.013 | 0 | - | - | - | - | - | - | - | - | - | - | - | - | 7.75 |
| K-Means | 0.257 | 0.22 | 0.346 | 0.087 | 0.18 | 0.113 | 0.133 | 0.046 | - | - | - | - | - | - | - | - | - | - | - | - | 6.85 |
| K-Medoids | 0.137 | 0.109 | 0.2 | 0.053 | 0.206 | 0.02 | 0 | 0 | - | - | - | - | - | - | - | - | - | - | - | - | 6.85 |
| NMF | 0.224 | 0.178 | 0.29 | 0.096 | 0.2 | 0.132 | 0.148 | 0.062 | 0.191 | 0.085 | 0.054 | 0.022 | 0.052 | 0.013 | 0.061 | 0.006 | 0.286 | 0.145 | 0.042 | 0.008 | 5.65 |
| Birch | 0.248 | 0.213 | 0.33 | 0.086 | 0.176 | 0.126 | 0.136 | 0.05 | - | - | - | - | - | - | - | - | - | - | - | - | 6.9 |
| NRP | 0.265 | 0.204 | 0.357 | 0.142 | 0.187 | 0.131 | 0.137 | 0.065 | 0.142 | 0.064 | 0.026 | 0 | 0.028 | 0.009 | 0.06 | 0.005 | 0.284 | 0.209 | 0.109 | 0.08 | 5.3 |
| PPR | 0.296 | 0.245 | 0.386 | 0.109 | 0.225 | 0.094 | 0.104 | 0.043 | - | - | - | - | - | - | - | - | - | - | - | - | 6.25 |
| SBC | 0.051 | 0.03 | 0.05 | 0.003 | 0.136 | 0.094 | 0.071 | 0.02 | - | - | - | - | - | - | - | - | - | - | - | - | 8.75 |
| SCC | 0.103 | 0.046 | 0.175 | 0.034 | 0.183 | 0.115 | 0.104 | 0.042 | - | - | - | - | - | - | - | - | - | - | - | - | 7.95 |
| biSBM-KL | - | - | - | - | 0.171 | 0.124 | 0.155 | 0.067 | - | - | - | - | - | - | - | - | - | - | - | - | 8.05 |
| biSBM-MCMC | 0.26 | 0.202 | 0.398 | 0.149 | 0.205 | 0.153 | 0.178 | 0.09 | - | - | - | - | - | - | - | - | - | - | - | - | 5.45 |
| HOPE | 0.354 | 0.308 | 0.453 | 0.204 | 0.228 | 0.16 | 0.177 | 0.087 | 0.24 | 0.087 | 0.083 | 0.031 | 0.129 | 0.043 | 0.164 | 0.033 | 0.405 | 0.191 | 0.225 | 0.181 | 2.85 |
| HOPE+ (FNEM) | 0.372 | 0.283 | 0.439 | 0.235 | 0.259 | 0.199 | 0.214 | 0.107 | 0.286 | 0.081 | 0.116 | 0.029 | 0.135 | 0.046 | 0.171 | 0.039 | 0.479 | 0.211 | 0.239 | 0.209 | 2.1 |
| HOPE+ (SNEM) | 0.391 | 0.277 | 0.448 | 0.248 | 0.277 | 0.17 | 0.208 | 0.124 | 0.288 | 0.094 | 0.12 | 0.053 | 0.169 | 0.047 | 0.18 | 0.059 | 0.507 | 0.221 | 0.25 | 0.234 | 1.25 |

**Baselines and Parameter Settings.** Table 3 summarizes the time complexities of our solutions and 13 competitors (the numbers of iterations in them are regarded as constants) evaluated in our experiments when adopted for $k$-BGC. Specifically, we compare HOPE, HOPE+ (FNEM), and HOPE+ (SNEM) against 5 unipartite graph clustering algorithms LE [39], Girvan–Newman [20], SC [55], NRP [64], and PPR [56], 4 data clustering methods including K-Means [24], K-Medoids [29], Birch [69], and NMF [61], as well as 4 BGC approaches, i.e., SBC [31], SCC [12], biSBM-KL [32] and biSBM-MCMC [67]. Note that in NRP and PPR, the clusters are obtained by applying K-Means over the node embedding vectors and PPR vectors they generate, respectively. The parameters of all competitors are set as suggested in their respective papers. Unless otherwise specified, we set $\alpha = 0.3$ and $\beta = 5k$ ($k$ is the number of clusters in $|\mathcal{U}|$) for HOPE and HOPE+. The number $T$ of iterations in HOPE+ (FNEM) and HOPE+ (SNEM) is set to 100.

**Evaluation Metrics.** We adopt 4 classic metrics to measure the clustering quality [60], including *Clustering Accuracy* (Acc), *F1 score* (F1), *Normalized Mutual Information* (NMI) [50], and *Adjusted Rand Index* (ARI) [27]. All metrics are calculated based on the ground-truth clustering labels and predicted clustering labels. Particularly, Acc, F1, and NMI range from 0 to 1.0, while ARI ranges from −0.5 to 1.0. For all of them, the higher values indicate better clustering quality. In terms of clustering efficiency and scalability, we report the running time in seconds (measured in wall-clock time) of each algorithm on each dataset, excluding the time for loading datasets and outputting clusters. A method is excluded if it cannot finish within 2 days.

### 5.2 Clustering Quality

Table 4 and Table 5 report the Acc, F1, NMI, ARI scores, as well as the average performance rankings of all methods on 5 small bipartite graph datasets (i.e., CORA, CiteSeer, BlogCatalog, Flickr, and PubMed) and 5 large datasets with more than a million edges, including CORA-F, LastFM (Asia), MIND, LastFM, and MAG. We report the best performance of HOPE, HOPE+ (FNEM), and HOPE+ (SNEM) on every bipartite graph data when varying $\beta$ from $2k$ to $8k$ and $\alpha$ from 0.1 to 0.9. The best, 2nd-best, and 3rd-best results among all methods are highlighted in blue, and darker shades indicate better scores. The best performance of all existing methods is underlined.

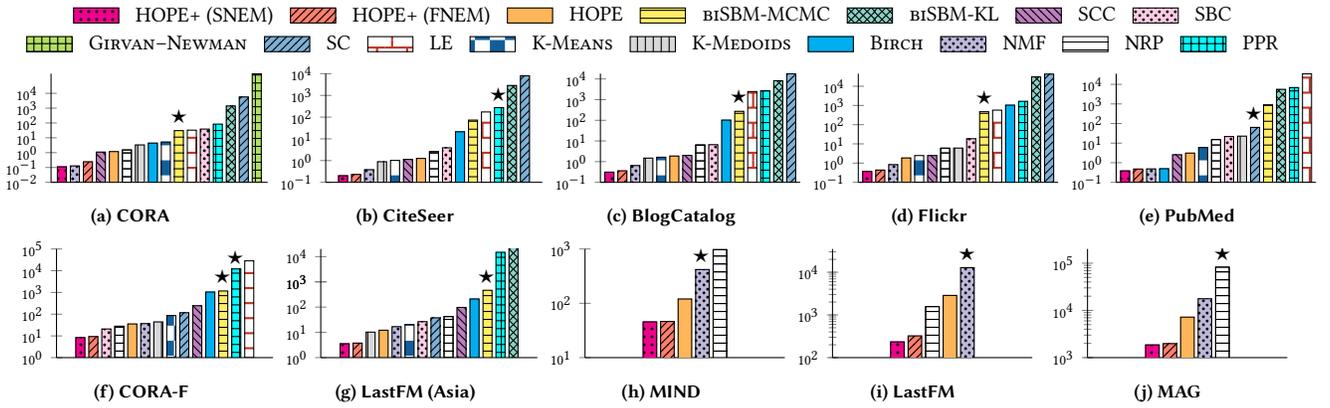

**Figure 7: Running time in seconds (★ marks the competitors with the best clustering quality in Tables 4 and 5).**

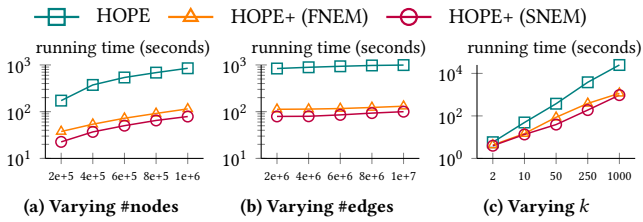

**Figure 8: Scalability tests.**

As shown in the last column of Tables 4 and 5, our methods, HOPE, HOPE+ (FNEM), and HOPE+ (SNEM), achieve the top-3 best average rank (smaller rank is better) on all datasets. Specifically, our proposed solutions considerably outperform all the competitors under most metrics on all 10 datasets. For instance, compared to existing methods, in Table 4, HOPE+ (SNEM) takes a lead by 12.7%, 5.5%, 13.4%, 3.5%, and 1.4% in terms of clustering accuracy (Acc) on the 5 small bipartite graphs, respectively. On the 5 large datasets in Table 5, HOPE+ (SNEM) outperforms existing approaches by a large margin of 9.5%, 5.2%, 9.7%, 11.7%, and 22.1% in terms of Acc, respectively. As for the other 3 metrics (*i.e.*, F1, NMI, and ARI), we can make qualitatively analogous observations, except on PubMed, where K-Means and SC obtain the highest NMI and ARI scores. The overall results in Table 4 and Table 5 validate the effectiveness of the proposed clustering objective via HOP vectors, which is then solved by the techniques developed in Sections 3 and 4.

In addition, in Tables 4 and 5, observe that HOPE+ (SNEM) has the best performance rank on all datasets, whereas HOPE and HOPE+ (FNEM) attain the second-best average performance rank on the small and large datasets, respectively. First, this indicates that HOPE inherits the local optima trap issue caused by the k-Means, and it is accentuated on large bipartite graphs, which is sidestepped by HOPE+ that does not rely on a new optimization framework presented in Section 4. Second, HOPE+ (SNEM) outperforms HOPE+ (FNEM) under most cases, which manifests that spectral norm is probably a better choice in Eq. (13) for our methods to handle $k$-BGC.

## 5.3 Efficiency and Scalability

Figure 7 depicts the respective running time of all methods required for clustering on the 10 bipartite graphs. The $y$-axis represents the running time (seconds) in the log-scale.

Observe that HOPE+ (SNEM) achieves the highest efficiency on all datasets, outperforming existing methods often by up to orders of magnitude. HOPE+ (FNEM) is also faster than all competitors, except on the smallest CORA. Particularly, on small bipartite graphs in Figure 7(a)-7(e), compared with the competitor with the best clustering quality in Table 4, *i.e.*, BISBM-MCMC, PPR, SCC, or K-Means, HOPE+ achieves over 270× speedup on CORA, CiteSeer, BlogCatalog, and Flickr, as well as more than 6.6× speedup on PubMed. As an example, on BlogCatalog, the best competitor BISBM-MCMC takes 271.8 seconds to complete, while both HOPE+ (SNEM) and HOPE+ (FNEM) need only around 0.32 seconds, attaining a significant speedup of 850×, respectively, while, as mentioned, the clustering quality of HOPE+ (SNEM) and HOPE+ (FNEM) is better than BISBM-MCMC. Compared with the fastest competitor NMF with inferior quality, HOPE+ (SNEM) is still consistently faster.

In Figure 7(f)-7(j) on large datasets, the efficiency improvement of our methods maintains or enlarges over existing methods, particularly on the three largest datasets MIND, LastFM, and MAG, where only our methods HOPE, HOPE+ with SNEM and FNEM, and competitors NRP, NMF survive to return within 2 days. Specifically, our solution HOPE+ (SNEM) is able to gain 9.2×, 54.7×, and 44.4× runtime speedup over the best viable competitor NMF or NRP as shown in Figure 7(h), 7(i), and 7(j), respectively. Notice that in spite of the comparable asymptotic complexities of NMF, SBC, and SCC as shown in Table 3, their efficiency significantly falls short compared to HOPE and HOPE+ due to numerous iterations required and/or time-consuming and complex operations involved in them.

The superior efficiency of HOPE+ demonstrates the power of the techniques proposed in Section 4 to significantly reduce computational costs, while achieving the state-of-the-art clustering quality as reported in Section 5.2. HOPE+ (SNEM) is consistently faster than HOPE+ (FNEM) since the former has lower time complexity as analyzed in Section 4.3. HOPE+ methods (Section 4) are more efficient than HOPE (Section 3) since HOPE+ avoids the expensive k-Means that needs many iterations to converge, via novel theoretical analysis and algorithm designs.

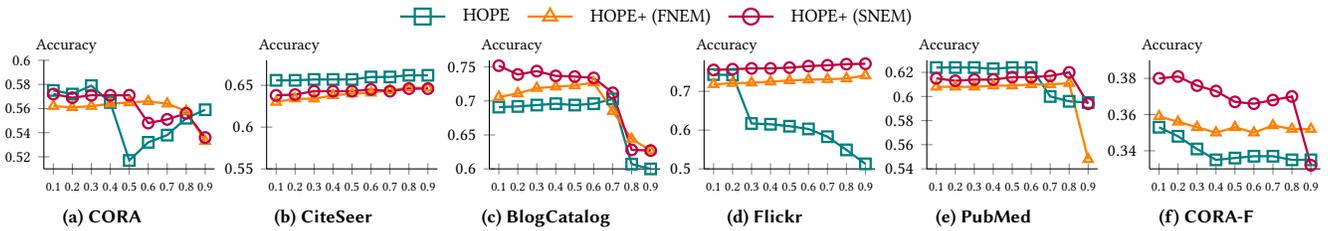

**Figure 9: Accuracy vs. $\alpha$.**

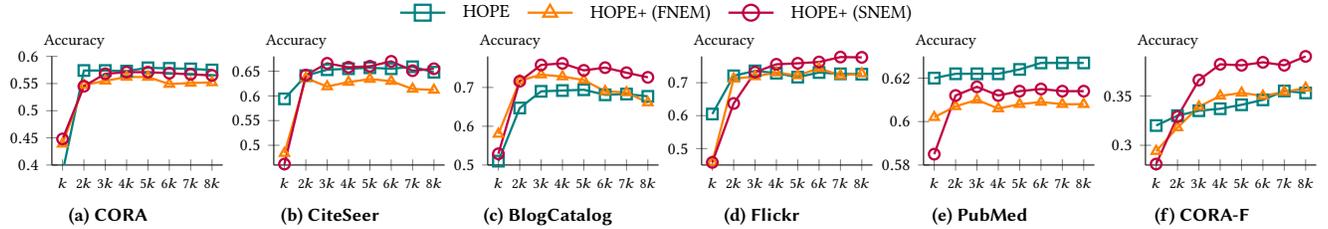

**Figure 10: Accuracy vs. $\beta$.**

Next, we verify the scalability of HOPE and HOPE+ on synthetic graphs of different sizes by the Erdős-Rényi random graph model [4]. Figure 8(a) shows the running time of HOPE, HOPE+ (FNEM), and HOPE+ (SNEM) on the random graphs generated by fixing the total number of vertices $|\mathcal{U}| + |\mathcal{V}| = 5 \times 10^5$ with $|\mathcal{U}| = |\mathcal{V}|$, and varying the number of edges $|\mathcal{E}|$ in $\{2 \times 10^6, 4 \times 10^6, 6 \times 10^6, 8 \times 10^6, 10^7\}$ when $k = 50$. Figure 8(b) displays the efficiency when varying the number of vertices $|\mathcal{U}| + |\mathcal{V}|$ in $\{2 \times 10^5, 4 \times 10^5, 6 \times 10^5, 8 \times 10^5, 10^6\}$, while fixing the number of edges $|\mathcal{E}| = 2 \times 10^6$ and $k = 50$. In Figure 8(c), we vary $k$ in $\{2, 10, 50, 250, 1000\}$ to evaluate the scalability, while fixing $|\mathcal{E}| = 2 \times 10^6$ and $|\mathcal{U}| + |\mathcal{V}| = 2 \times 10^5$. Observe that the runtime of our methods increases steadily in proportion to the number of vertices and the number $k$ of clusters, but grows modestly as the number of edges increases.

## 5.4 Parameter Analysis

We empirically study the effects of the input parameters of HOPE and HOPE+, including the random walk decay factor $\alpha$ and the dimensionality $\beta$.

**Varying $\alpha$.** Figure 9 illustrates the Acc scores of HOPE, HOPE+ (FNEM), and HOPE+ (SNEM) on 6 datasets, including CORA, CiteSeer, BlogCatalog, Flickr, PubMed, and CORA-F, when $\alpha$ is varied from 0.1 to 0.9 with step size 0.1. The F1, NMI, and ARI results are qualitatively similar, and thus, are omitted here for the lack of space. From Figure 9, we can see that, on CORA, BlogCatalog, and PubMed, the clustering accuracy scores achieved by HOPE and HOPE+ are stable at first, and then decrease remarkably when $\alpha$ roughly is beyond 0.5. Additionally, the clustering performance of all three methods witnesses a significant drop on CORA-F after $\alpha$ exceeds 0.2. Recall that in Eq. (5), a larger $\alpha$ results in higher weights (*i.e.*, $(1 - \alpha)\alpha^\lambda$) to distant vertices in **H**. Therefore, it can be concluded that on CORA, BlogCatalog, PubMed, and CORA-F, the affinities of the given vertex $u_i$ and its proximal vertices play a more important role in the clustering. Interestingly, on CiteSeer, the performance of both HOPE and HOPE+ increases slightly with

$\alpha$, implying the consideration of more far-reaching vertices benefits the clustering. A similar observation can be made on Flickr for HOPE+, while the accuracy scores of HOPE slump by more than 10% when $\alpha > 0.2$. Further, Figure 9 reveals that HOPE+ yields better stability over HOPE, particularly on CORA and Flickr, when varying $\alpha$.

**Varying $\beta$.** Figure 10 reports the clustering accuracy of our methods when varying $\beta$ from $k$ to $8k$ ($k$ is the number of clusters per dataset). Observe that, on most datasets, the clustering accuracy scores attained by HOPE, HOPE+ (FNEM), and HOPE+ (SNEM) initially experience a rapid increase when $\beta$ is varied from $k$ to $3k$, followed by remaining stable with slight fluctuations when $\beta \geq 3k$. To explain, recall that $\beta$ is used as the dimensionality of the low-rank matrix **X** to approximate **H**. Intuitively, a larger $\beta$ gives a more accurate low-rank approximation of **H**, thereby leading to higher result quality in $k$-BGC. Furthermore, Figure 11 presents the running time in seconds of HOPE, HOPE+ (FNEM), and HOPE+ (SNEM) when $\beta$ is varied from $k$ to $8k$ on 3 representative large datasets CORA-F, MIND, and LastFM. The runtime of HOPE and HOPE+ is roughly proportional to $\beta$, coinciding with our asymptotic analysis of HOPE and HOPE+ in Sections 3 and 4.4, respectively.

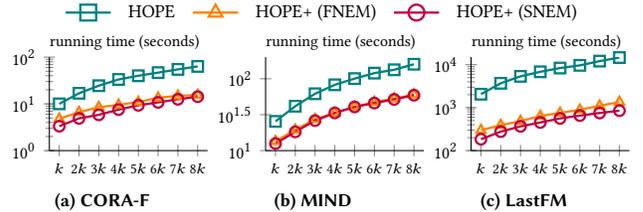

**Figure 11: Running time vs. $\beta$.**

## 6 RELATED WORK

This section reviews existing studies germane to the $k$-BGC problem in this paper. A common methodology for $k$-BGC is to first project a bipartite graph $\mathcal{G}$ into a unipartite graph $\mathcal{G}^*$, *i.e.*, a graph composed of a single type of vertices $\mathcal{U}$, by connecting two vertices from $\mathcal{U}$ if they share common neighborhoods in $\mathcal{G}$. After that, classic graph clustering algorithms can be applied naturally to $\mathcal{G}^*$. A variety of methods [70, 71] have been proposed for weighting the connections in $\mathcal{G}^*$. Melamed [36] proposed a dual-projection method, which extracts clusters of two one-mode graphs based on the two vertex sets in $\mathcal{G}$ independently and then combines the solutions such that the within-community ties are maximized. CoмSim [52] detects clusters of one single type of vertex using a similarity measure of common neighbors and maximizes vertices' similarity by identifying cycles of connections. In spite of also relying on projected graphs, our methods differ from previous projection-based methods. To be more precise, our solutions make use of the high-order affinities between target vertex set $\mathcal{U}$ and the vertices in the proposed weighted projected graph $\mathcal{G}_\mathcal{V}$ for dividing $\mathcal{U}$ into clusters, whereas prior methods partition vertices in $\mathcal{G}_\mathcal{V}$ based on their direct connections in $\mathcal{G}_\mathcal{V}$.

Recent $k$-BGC solutions propose to simultaneously group vertices in each vertex set of $\mathcal{G}$ into clusters. Dhillon [12] extended spectral graph partitioning [44] to bipartite graphs. SBC algorithm [31] finds a blockwise-constant checkerboard matrix as a good approximation of a bipartite graph $\mathcal{G}$ for clustering. The state-of-the-art solutions to $k$-BGC are based on statistical models. biSBM-KL [32] includes a degree-corrected bipartite stochastic block model for inferring clusters and maximizes the likelihood function using the Kernighan-Lin algorithm [30]. In follow-up work, Yen and Larremore [67] further developed biSBM-MCMC, which adopts a Markov Chain Monte-Carlo sampler for fast optimization. Our methods outperform these approaches, as validated by our experiments, due to our novel designs that efficiently and effectively exploit high-order relationships between vertices.

Instead of assuming the number $k$ of clusters to be known a priori, another line of research is devoted to discovering $k$ automatically during clustering and various techniques have been proposed. Lehmann et al. [33] presented biGclique for detecting biclique clusters in bipartite graphs by extending the $k$-clique community [41] to bipartite graphs. Considerable efforts have been devoted to extending the notion of modularity [40] in unipartite graphs to bipartite graphs as bi-modularity for jointly clustering two types of vertices based on bi-modularity maximization [2, 23]. Guimera et al. [23] use the cumulative deviation from the random expectation of the number of neighbors shared by vertices in the same cluster to define bi-modularity. Another bi-modularity objective is defined in [2, 3] with the consideration of connectivity of both vertex sets in a bipartite graph, and a BRIM algorithm is presented to solve the objective. Subsequent studies [13, 35, 37, 42, 51, 68] propose more bi-modularity definitions as well as related optimization algorithms. However, these modularity-based methods suffer from the *resolution limit* issue [18] due to failing to accurately detect small-sized clusters with high modularity scores.

In addition, classic graph clustering algorithms [17] and standard data clustering algorithms can be applied to $k$-BGC by considering bipartite graphs as general graphs and data matrices, respectively. Since they are not specially catered for bipartite graphs, these methods cannot capture the unique characteristics of bipartite structures, resulting in a mediocre performance, as shown in Section 5.

## 7 CONCLUSION

In this paper, we propose efficient solutions, HOPE and HOPE+ for high-quality $k$-BGC. HOPE achieves high scalability and clustering quality by formulating the $k$-BGC as an optimization problem based on our proposed HOP vectors and an efficient low-rank approximation of the HOP vectors. HOPE+ further improves over HOPE in terms of both empirical efficiency and clustering accuracy through a novel problem transformation and a carefully-crafted two-stage optimization framework, as demonstrated by our extensive experiments. In the future, we intend to extend HOPE and HOPE+ to handle bipartite graphs with vertex attributes.

## ACKNOWLEDGMENTS

Renchi Yang is supported by the NSFC YSF grant (No. 62302414) and Hong Kong RGC ECS grant (No. 22202623). Jieming Shi is supported by Hong Kong RGC ECS (No. 25201221) and NSFC 62202404.

## A PROOFS

**Proof of Lemma 2.2.** First, we define a matrix $\Omega \in \mathbb{R}^{|\mathcal{V}| \times |\mathcal{V}|}$ where $\Omega_{j,l} = \sum_{u_i \in \mathcal{U}} p(v_j, u_i) \cdot p(u_i, v_l)$. It can be seen that $\Omega$ is a non-negative row-stochastic matrix, i.e., $\sum_{v_l \in \mathcal{V}} \Omega_{j,l} = 1$. As per [1] and the definition of $\Omega$, there exists a probability distribution $\pi^{(j)}$, such that $\lim_{\lambda \to \infty} \Omega_j^\lambda = \pi$ for $1 \le j \le |\mathcal{V}|$.

Next, let $\Delta$ be a $|\mathcal{V}| \times |\mathcal{V}|$ diagonal matrix wherein $\Delta_{j,j}$ equals to $ws(v_j) = \sum_{u_h \in \mathcal{U}} w(v_j, u_h)$. By Eq. (1) and (4), we can derive

$$\mathbf{Q}_j \cdot \mathbf{Q}_l = \sum_{u_i \in \mathcal{U}} \sqrt{p(v_j, u_i) \cdot p(u_i, v_j)} \cdot \sqrt{p(v_l, u_i) \cdot p(u_i, v_l)}$$
$$= \sum_{u_i \in \mathcal{U}} \frac{w(v_j, u_i)}{\sqrt{ws(v_j)}} \cdot \frac{p(u_i, v_l)}{\sqrt{ws(v_l)}} = \Delta_{j,j}^{\frac{1}{2}} \cdot \Omega_{j,l} \cdot \Delta_{l,l}^{-\frac{1}{2}},$$

which implies that $\mathbf{Q}\mathbf{Q}^\top = \Delta^{\frac{1}{2}} \Omega \Delta^{-\frac{1}{2}}$. Thus, each term in Eq. (5) satisfies $\mathbf{P}(\mathbf{Q}\mathbf{Q}^\top)^\lambda = \mathbf{P}\Delta^{\frac{1}{2}} \Omega^\lambda \Delta^{-\frac{1}{2}}$. Since $\lim_{\lambda \to \infty} \Omega_j^\lambda = \pi^{(j)}$ $\forall 1 \le j \le |\mathcal{V}|$, each row of $\mathbf{P}(\mathbf{Q}\mathbf{Q}^\top)^\lambda$ will converge to a vector when $\lambda \to \infty$. In addition, $\Omega$ is a non-negative row-stochastic matrix, then $\Omega^\lambda$ is also a non-negative row-stochastic matrix, meaning that each entry in $\Omega$ is not greater than 1. Therefore, $(\mathbf{P}(\mathbf{Q}\mathbf{Q}^\top)^\lambda)_{i,j} = (\mathbf{P}\Delta^{\frac{1}{2}} \Omega^\lambda \Delta^{-\frac{1}{2}})_{i,j} \le \mathbf{P}_{i,j}$ holds for any $\lambda \ge 0$, $u_i \in \mathcal{U}$ and $v_j \in \mathcal{V}$, which further leads to

$$\mathbf{F}_{i,j} = \sum_{\lambda=0}^{\infty} (1-\alpha)\alpha^\lambda \cdot (\mathbf{P}(\mathbf{Q}\mathbf{Q}^\top)^\lambda)_{i,j} \le \sum_{\lambda=0}^{\infty} (1-\alpha)\alpha^\lambda \cdot \mathbf{P}_{i,j}$$
$$\le \mathbf{P}_{i,j} \cdot \sum_{\lambda=0}^{\infty} (1-\alpha)\alpha^\lambda = \mathbf{P}_{i,j} \le 1.$$

Since $\mathbf{P}(\mathbf{Q}\mathbf{Q}^\top)^\lambda$ is non-negative, $\mathbf{F}_{i,j} \ge 0$, completing the proof. □

**Proof of Lemma 3.1.** By the definition of SVD, $\mathbf{Q} = \mathbf{U}\Sigma\mathbf{V}^\top$ and right singular vectors $\mathbf{V}$ have orthogonal columns, i.e., $\mathbf{V}^\top\mathbf{V} = \mathbf{I}$. Further, we have $\mathbf{Q}\mathbf{Q}^\top = \mathbf{U}\Sigma^2\mathbf{U}^\top$. Notice that the left singular vectors $\mathbf{U}$ also have orthogonal columns, i.e., $\mathbf{U}^\top\mathbf{U} = \mathbf{I}$. Hence,

$$\sum_{\lambda=0}^{\infty} (1-\alpha)\alpha^\lambda (\mathbf{Q}\mathbf{Q}^\top)^\lambda = \sum_{\lambda=0}^{\infty} (1-\alpha)\alpha^\lambda (\mathbf{U}\Sigma^2\mathbf{U}^\top)^\lambda$$
$$= \mathbf{U} \sum_{\lambda=0}^{\infty} (1-\alpha)\alpha^\lambda \Sigma^{2\lambda} \mathbf{U}^\top$$
$$= \mathbf{U} \left( \frac{(1-\alpha)(1-(\alpha\Sigma^2)^\infty)}{1-\alpha\Sigma^2} \right) \mathbf{U}^\top \quad (16)$$

Continuing forth, we prove that the largest singular value of $\mathbf{Q}$ is not greater than 1, namely $\Sigma_{1,1} \le 1$. Firstly, as per Lemma 4.3, the largest singular value of $\mathbf{Q}$ is equal to the largest eigenvalue of $\mathbf{Q}\mathbf{Q}^\top$. Thus, if we can prove that the largest eigenvalue of $\mathbf{Q}\mathbf{Q}^\top$ is less than or equal to 1, then $\Sigma_{1,1} \leq 1$ holds. For any $v_i \in \mathcal{V}$, let $z_i(u_l) = \sqrt{p(v_i, u_l) \cdot p(u_l, v_i)}$. By Eq. (2), we can write $(\mathbf{QQ}^\top)_{i,j}$ as

$$(\mathbf{QQ}^\top)_{i,j} = \sum_{u_l \in \mathcal{U}} \sqrt{p(v_j, u_l) \cdot p(u_l, v_j)} \cdot \sqrt{p(v_i, u_l) \cdot p(u_l, v_i)}$$
$$= \sum_{u_l \in \mathcal{U}} z_j(u_l) \cdot z_i(u_l)$$

Also, on the basis of Eq. (1), it is easy to prove that

$$\sum_{u_l \in \mathcal{U}} z_i(u_l)^2 = \sum_{u_l \in \mathcal{U}} p(v_i, u_l) \cdot p(u_l, v_i) \leq \max_{u_l \in \mathcal{U}} p(v_i, v_i) \sum_{u_l \in \mathcal{U}} p(v_i, u_l) \leq 1$$

Then, for any vector $\mathbf{y} \in \mathbb{R}^{|\mathcal{V}|}$, the following inequality holds:

$$\mathbf{y} \cdot (\mathbf{I} - \mathbf{QQ}^\top) \cdot \mathbf{y}^\top = \sum_{v_i \in \mathcal{V}} y_i^2 - 2 \sum_{v_i, v_j \in \mathcal{V}} y_i \cdot (\mathbf{QQ}^\top)_{i,j} \cdot y_j$$
$$= \sum_{v_i \in \mathcal{V}} y_i^2 - 2 \sum_{v_i, v_j \in \mathcal{V}} y_i \cdot y_j \cdot \sum_{u_l \in \mathcal{U}} z_j(u_l) \cdot z_i(u_l)$$
$$= \sum_{v_i \in \mathcal{V}} y_i^2 \cdot \sum_{u_l \in \mathcal{U}} z_i(u_l)^2 + \sum_{v_i \in \mathcal{V}} y_i^2 \cdot (1 - \sum_{u_l \in \mathcal{U}} z_i(u_l)^2)$$
$$- 2 \sum_{v_i, v_j \in \mathcal{V}} y_i \cdot y_j \cdot \sum_{u_l \in \mathcal{U}} z_j(u_l) \cdot z_i(u_l)$$
$$= \sum_{v_i \in \mathcal{V}} \sum_{u_l \in \mathcal{U}} (y_i \cdot z_i(u_l) - y_j \cdot z_j(u_l))^2 + \sum_{v_i \in \mathcal{V}} y_i^2 \cdot (1 - \sum_{u_l \in \mathcal{U}} z_i(u_l))^2 \geq 0,$$

which indicates that $\mathbf{QQ}^\top$ is positive semidefinite, and hence, gives us the following Rayleigh quotient

$$\mathbf{y} \cdot (\mathbf{I} - \mathbf{QQ}^\top) \cdot \mathbf{y}^\top \geq 0 \Longrightarrow 1 \geq \frac{\mathbf{y} \cdot \mathbf{QQ}^\top \cdot \mathbf{y}^\top}{\mathbf{y}\mathbf{y}^\top}.$$

The above inequality implies that the largest eigenvalue of $\mathbf{QQ}^\top$ is less than or equal to 1, and accordingly $\Sigma_{1,1} \leq 1$. Furthermore, we can derive that $(\alpha \Sigma^2)^\infty = 0$ and plugging it into Eq. (16) yields

$$\sum_{\lambda=0}^{\infty} (1-\alpha) \alpha^\lambda (\mathbf{QQ}^\top)^\lambda = \mathbf{U} \frac{1-\alpha}{1-\alpha \Sigma^2} \mathbf{U}^\top. \tag{17}$$

The lemma is proved. □

**Proof of Theorem 3.2.** Since the $L_2$ norm of any HOP vector $\mathbf{H}_i$ is 1 (E.q (6)), we then derive that

$$\|\mathbf{H}_i - \mathbf{H}_j\|_2^2 = \sum_{l=1}^{|\mathcal{V}|} (\mathbf{H}_{i,l} - \mathbf{H}_{j,l})^2 = \sum_{l=1}^{|\mathcal{V}|} \mathbf{H}_{i,l}^2 + \mathbf{H}_{j,l}^2 - 2\mathbf{H}_{i,l} \cdot \mathbf{H}_{j,l}$$
$$= \sum_{l=1}^{|\mathcal{V}|} \mathbf{H}_{i,l}^2 + \sum_{l=1}^{|\mathcal{V}|} \mathbf{H}_{j,l}^2 + 2 \sum_{l=1}^{|\mathcal{V}|} \mathbf{H}_{i,l} \cdot \mathbf{H}_{j,l}$$
$$= \|\mathbf{H}_i\|_2^2 + \|\mathbf{H}_j\|_2^2 - 2\mathbf{H}_i \cdot \mathbf{H}_j = 2(1 - \mathbf{H}_i \cdot \mathbf{H}_j).$$

Likewise, $\|\mathbf{X}_i - \mathbf{X}_j\|_2^2 = 2(1 - \mathbf{X}_i \cdot \mathbf{X}_j)$ since each row of $\mathbf{X}$ has a unit $L_2$ norm (Line 4 in Algorithm 1). In addition, by Eq. (8) and the fact that $\|\widehat{\mathbf{X}}_i\|_2 = \sqrt{\widehat{\mathbf{X}}_i \cdot \widehat{\mathbf{X}}_i}$, we can get $\mathbf{X}_i \cdot \mathbf{X}_j = \frac{\widehat{\mathbf{X}}_i \cdot \widehat{\mathbf{X}}_j}{\|\widehat{\mathbf{X}}_i\|_2 \cdot \|\widehat{\mathbf{X}}_j\|_2} = \frac{\widehat{\mathbf{X}}_i \cdot \widehat{\mathbf{X}}_j}{\sqrt{\widehat{\mathbf{X}}_i \widehat{\mathbf{X}}_i} \cdot \sqrt{\widehat{\mathbf{X}}_j \widehat{\mathbf{X}}_j}}$. By Eq. (8), Section 3, and Eq. (17), $\widehat{\mathbf{X}}$ satisfies

$$\widehat{\mathbf{X}} \cdot \widehat{\mathbf{X}}^\top = \mathbf{P}\mathbf{U}(\frac{1-\alpha}{1-\alpha \cdot \Sigma^2})^2 \mathbf{U}^\top \mathbf{P}^\top, \tag{18}$$

and particularly, $\mathbf{F}\mathbf{F}^\top = \mathbf{P}\overline{\mathbf{U}}(\frac{1-\alpha}{1-\alpha \cdot \overline{\Sigma}^2})^2 \overline{\mathbf{U}}^\top \mathbf{P}^\top$, where $\overline{\mathbf{U}}$ and $\overline{\Sigma}$ are the full left singular vectors and singular values of $\mathbf{Q}$, respectively. By the fact that every element in $\mathbf{P}$ is less than 1, we have

$$\|\mathbf{F}\mathbf{F}^\top - \widehat{\mathbf{X}}\widehat{\mathbf{X}}^\top\|_2 = \|\mathbf{P}\overline{\mathbf{U}}(\frac{1-\alpha}{1-\alpha \cdot \overline{\Sigma}^2})^2 \overline{\mathbf{U}}^\top \mathbf{P}^\top - \mathbf{P}\mathbf{U}(\frac{1-\alpha}{1-\alpha \cdot \Sigma^2})^2 \mathbf{U}^\top \mathbf{P}^\top\|_2$$
$$\leq \|\overline{\mathbf{U}}(\frac{1-\alpha}{1-\alpha \overline{\Sigma}^2})^2 \overline{\mathbf{U}}^\top - \mathbf{U}(\frac{1-\alpha}{1-\alpha \Sigma^2})^2 \mathbf{U}^\top\|_2. \tag{19}$$

**Theorem A.1** (Eckart–Young Theorem [21]). *Suppose that $\mathbf{M}_k \in \mathbb{R}^{n \times k}$ is the rank-$k$ approximation to $\mathbf{M} \in \mathbb{R}^{n \times n}$ obtained by exact SVD, then $\min_{rank(\widetilde{\mathbf{M}}) \leq k} \|\mathbf{M} - \widehat{\mathbf{M}}\|_2 = \|\mathbf{M} - \mathbf{M}_k\|_2 = \sigma_{k+1}$, where $\sigma_i$ represents the $i$-th largest singular value of $\mathbf{M}$.*

As per the definitions, $\mathbf{U}(\frac{1-\alpha}{1-\alpha \cdot \Sigma^2})^2 \mathbf{U}^\top$ is the rank-$k$ approximation of $\overline{\mathbf{U}}(\frac{1-\alpha}{1-\alpha \cdot \overline{\Sigma}^2})^2 \overline{\mathbf{U}}^\top$. By Theorem A.1, Inequality (19) becomes

$$\|\mathbf{F}\mathbf{F}^\top - \widehat{\mathbf{X}}\widehat{\mathbf{X}}^\top\|_2 \leq \left\|\overline{\mathbf{U}}(\frac{1-\alpha}{1-\alpha \cdot \overline{\Sigma}^2})^2 \overline{\mathbf{U}}^\top - \mathbf{U}(\frac{1-\alpha}{1-\alpha \cdot \Sigma^2})^2 \mathbf{U}^\top\right\|_2 = \sigma,$$

where $\sigma = (\frac{1-\alpha}{1-\alpha \cdot \overline{\Sigma}_{\beta+1,\beta+1}^2})^2$ and $\overline{\Sigma}_{\beta+1,\beta+1}$ is the $(\beta + 1)$-th largest singular value of $\mathbf{Q}$. Since for any matrix $\mathbf{M}$, $\|\mathbf{M}\|_{max} \leq \|\mathbf{M}\|_2$, we have $\|\mathbf{F}\mathbf{F}^\top - \widehat{\mathbf{X}}\widehat{\mathbf{X}}^\top\|_{max} \leq \sigma$, leading to $|\widehat{\mathbf{X}}_i \cdot \widehat{\mathbf{X}}_j - \mathbf{F}_i \cdot \mathbf{F}_j| \leq \sigma$, $\forall u_l \in \mathcal{U}$. Recall that $\mathbf{H}_i = \frac{\mathbf{F}_i}{\|\mathbf{F}_i\|_2}$ in Eq. (6). Consequently, we have

$$\frac{\mathbf{H}_i \cdot \mathbf{H}_j \cdot \|\mathbf{F}_i\|_2 \cdot \|\mathbf{F}_j\|_2 - \sigma}{\sqrt{\|\mathbf{F}_i\|_2^2 + \sigma} \cdot \sqrt{\|\mathbf{F}_j\|_2^2 + \sigma}} \leq \mathbf{X}_i \cdot \mathbf{X}_j \leq \frac{\mathbf{H}_i \cdot \mathbf{H}_j \cdot \|\mathbf{F}_i\|_2 \cdot \|\mathbf{F}_j\|_2 + \sigma}{\sqrt{\|\mathbf{F}_i\|_2^2 - \sigma} \cdot \sqrt{\|\mathbf{F}_j\|_2^2 - \sigma}} \tag{20}$$

Note that $\|\mathbf{X}_i - \mathbf{X}_j\|_2^2 - \|\mathbf{H}_i - \mathbf{H}_j\|_2^2 = 2(\mathbf{H}_i \cdot \mathbf{H}_j - \mathbf{X}_i \cdot \mathbf{X}_j)$. Plugging Eq. (20) into the above equation completes the proof. □

**Proof of Lemma 4.4.** First, by the definition of matrix Frobenius norm and matrix trace property,

$$\min_{\mathbf{T}} \|\mathbf{L}\mathbf{T} - \mathbf{C}\|_F^2 = \min_{\mathbf{T}} \langle \mathbf{L}\mathbf{T} - \mathbf{C}, \mathbf{L}\mathbf{T} - \mathbf{C} \rangle_F$$
$$= \min_{\mathbf{T}} \|\mathbf{L}\mathbf{T}\|_F^2 + \|\mathbf{C}\|_F^2 - 2\langle \mathbf{L}\mathbf{T}, \mathbf{C} \rangle$$
$$= \min_{\mathbf{T}} \|\mathbf{L}\|_F^2 + \|\mathbf{C}\|_F^2 - 2\langle \mathbf{L}\mathbf{T}, \mathbf{C} \rangle_F$$
$$= 2(1 - \min_{\mathbf{T}} Tr(\mathbf{L}\mathbf{T}\mathbf{C}^\top)) = 2(1 - \min_{\mathbf{T}} Tr(\mathbf{T}(\mathbf{L}^\top \mathbf{C})^\top).$$

Suppose that $\mathbf{\Phi}\mathbf{\Sigma}\mathbf{\Psi}^\top$ is the exact full SVD of $\mathbf{L}^\top \mathbf{C}$, where the left and right singular vectors satisfy $\mathbf{\Phi}^\top \mathbf{\Phi} = \mathbf{\Psi}^\top \mathbf{\Psi} = \mathbf{I}$. Further, by the cyclic property of matrix trace and the above equation, $\min_{\mathbf{T}} \|\mathbf{L}\mathbf{T} - \mathbf{C}\|_F$ is equivalent to

$$\max_{\mathbf{T}} Tr(\mathbf{T} \cdot (\mathbf{L}^\top \mathbf{C})^\top) = \max_{\mathbf{T}} Tr(\mathbf{T}\mathbf{\Psi}\mathbf{\Sigma}\mathbf{\Phi}^\top) = \max_{\mathbf{T}} Tr(\mathbf{\Phi}^\top \mathbf{T}\mathbf{\Psi}\mathbf{\Sigma})$$
$$= \max_{\mathbf{T}} \sum_{i=1}^k (\mathbf{\Phi}^\top \mathbf{T}\mathbf{\Psi})_{i,i} \cdot \Sigma_{i,i}$$

Observe that the rows of $\mathbf{\Phi}^\top \mathbf{T}\mathbf{\Psi}$ are orthogonal, namely $\mathbf{\Phi}^\top \mathbf{T}\mathbf{\Psi} \cdot (\mathbf{\Phi}^\top \mathbf{T}\mathbf{\Psi})^\top = \mathbf{I}$, meaning that $-1 \leq (\mathbf{\Phi}^\top \mathbf{T}\mathbf{\Psi})_{i,j} \leq 1$ $\forall 1 \leq i, j \leq k$. Therefore, we have $\sum_{i=1}^k (\mathbf{\Phi}^\top \mathbf{T}\mathbf{\Psi})_{i,i} \cdot \Sigma_{i,i} \leq \sum_{i=1}^k \Sigma_{i,i}$. In particular, the upper bound $\sum_{i=1}^k \Sigma_{i,i}$ is attained when $\mathbf{\Phi}^\top \mathbf{T}\mathbf{\Psi} = \mathbf{I}$, implying that $\mathbf{T} = \mathbf{\Phi}\mathbf{\Psi}^\top$. Thus, the lemma holds. □

**Proof of Lemma 4.5.** Suppose that $\mathbf{T}^*$ is the optimal solution to the objective function in Eq. (15). In the first place, it must hold that $\|\mathbf{L}\mathbf{T}^* - \mathbf{C}\|_2 \leq \|\mathbf{L}\mathbf{L}^\top \mathbf{C} - \mathbf{C}\|_2$ since $\mathbf{T}^*$ is the minimizer. On the other hand, by the Pythagorean theorem [57], for any vector $\mathbf{v}$,

$$\|(\mathbf{L}\mathbf{T}^* - \mathbf{C})\mathbf{v}\|_2^2 = \|(\mathbf{L}\mathbf{L}^\top \mathbf{C} - \mathbf{C})\mathbf{v}\|_2^2 + \|(\mathbf{L}\mathbf{T}^* - \mathbf{L}\mathbf{L}^\top \mathbf{C})\mathbf{v}\|_2^2$$
$$\geq \|(\mathbf{L}\mathbf{L}^\top \mathbf{C} - \mathbf{C})\mathbf{v}\|_2^2,$$

and hence $\|\mathbf{L}\mathbf{T}^* - \mathbf{C}\|_2 \geq \|\mathbf{L}\mathbf{L}^\top \mathbf{C} - \mathbf{C}\|_2$. As a consequence, $\|\mathbf{L}\mathbf{T}^* - \mathbf{C}\|_2 = \|\mathbf{L}\mathbf{L}^\top \mathbf{C} - \mathbf{C}\|_2$ and

$$\mathbf{L}\mathbf{T}^* = \mathbf{L}\mathbf{L}^\top \mathbf{C} \tag{21}$$

Moreover, recall that $\mathbf{L}$ is the $k$-largest eigenvectors, which satisfies $\mathbf{L}^\top \mathbf{L} = \mathbf{I}$. Accordingly, as per Eq. (21), $\mathbf{T}^* = \mathbf{L}^\top \mathbf{L}\mathbf{T}^* = \mathbf{L}^\top \mathbf{L}\mathbf{L}^\top \mathbf{C} = \mathbf{L}^\top \mathbf{C}$, signifying that the optimal such $\mathbf{T}$ in Eq. (15) is $\mathbf{L}^\top \mathbf{C}$. The proof is completed. □

# REFERENCES

[1] Jac M Anthonisse and Henk Tijms. 1977. Exponential convergence of products of stochastic matrices. *J. Math. Anal. Appl.* 59, 2 (1977), 360–364.

[2] Michael J Barber. 2007. Modularity and community detection in bipartite networks. *Physical Review E* 76, 6 (2007), 066102.

[3] Michael J Barber, Margarida Faria, Ludwig Streit, and Oleg Strogan. 2008. Searching for communities in bipartite networks. In *AIP Conference Proceedings*, Vol. 1021. American Institute of Physics, 171–182.

[4] Vladimir Batagelj and Ulrik Brandes. 2005. Efficient generation of large random networks. *Physical Review E* 71, 3 (2005), 036113.

[5] Aleksandar Bojchevski and Stephan Günnemann. 2018. Deep Gaussian Embedding of Graphs: Unsupervised Inductive Learning via Ranking. In *International Conference on Learning Representations*. https://openreview.net/forum?id=r1ZdKJ-0W

[6] Aleksandar Bojchevski, Johannes Klicpera, Bryan Perozzi, Amol Kapoor, Martin Blais, Benedek Rózemberczki, Michal Lukasik, and Stephan Günnemann. 2020. Scaling Graph Neural Networks with Approximate PageRank. In *SIGKDD*.

[7] Gabriel Budel and Piet Van Mieghem. 2020. Detecting the number of clusters in a network. *Journal of Complex Networks* 8, 6 (2020), cnaa047.

[8] Paola Gabriela Pesántez Cabrera. 2018. *Bipartite Network Community Detection: Algorithms and Applications*. Washington State University.

[9] O. Celma. 2010. *Music Recommendation and Discovery in the Long Tail*. Springer.

[10] Yizong Cheng and George M Church. 2000. Biclustering of expression data.. In *Ismb*, Vol. 8. 93–103.

[11] Wenyuan Dai, Gui-Rong Xue, Qiang Yang, and Yong Yu. 2007. Co-clustering based classification for out-of-domain documents. In *Proceedings of the 13th ACM SIGKDD international conference on Knowledge discovery and data mining*. 210–219.

[12] Inderjit S Dhillon. 2001. Co-clustering documents and words using bipartite spectral graph partitioning. In *Proceedings of the seventh ACM SIGKDD international conference on Knowledge discovery and data mining*. 269–274.

[13] Carsten F Dormann and Rouven Strauss. 2014. A method for detecting modules in quantitative bipartite networks. *Methods in Ecology and Evolution* 5, 1 (2014), 90–98.

[14] Jennifer A Dunne, Richard J Williams, and Neo D Martinez. 2002. Network structure and biodiversity loss in food webs: robustness increases with connectance. *Ecology letters* 5, 4 (2002), 558–567.

[15] Ky Fan. 1949. On a theorem of Weyl concerning eigenvalues of linear transformations I. *Proceedings of the National Academy of Sciences of the United States of America* 35, 11 (1949), 652.

[16] Yiling Chen Frederico Fonseca. 2003. A bipartite graph co-clustering approach to ontology mapping. In *Proceedings of the Workshop on Semantic Web Technologies for Searching and Retrieving Scientific Data. Colocated with the Second International Semantic Web Conference (ISWC-03), CEUR-WS. org.*

[17] Santo Fortunato. 2010. Community detection in graphs. *Physics reports* 486, 3-5 (2010), 75–174.

[18] Santo Fortunato and Marc Barthelemy. 2007. Resolution limit in community detection. *Proceedings of the national academy of sciences* 104, 1 (2007), 36–41.

[19] Lise Getoor. 2005. Link-based classification. In *Advanced methods for knowledge discovery from complex data*. Springer, 189–207.

[20] Michelle Girvan and Mark EJ Newman. 2002. Community structure in social and biological networks. *Proceedings of the national academy of sciences* 99, 12 (2002), 7821–7826.

[21] Gene H Golub and Charles F Van Loan. 1996. *Matrix computations*. Johns Hopkins University, Press (1996).

[22] John C Gower and Garmt B Dijksterhuis. 2004. *Procrustes problems*. Vol. 30. OUP Oxford.

[23] Roger Guimera, Marta Sales-Pardo, and Luis A Nunes Amaral. 2007. Module identification in bipartite and directed networks. *Physical Review E* 76, 3 (2007), 036102.

[24] John A Hartigan and Manchek A Wong. 1979. Algorithm AS 136: A k-means clustering algorithm. *Journal of the royal statistical society. series c (applied statistics)* 28, 1 (1979), 100–108.

[25] Reinhard Heckel, Michail Vlachos, Thomas Parnell, and Celestine Dünner. 2017. Scalable and interpretable product recommendations via overlapping co-clustering. In *2017 IEEE 33rd International Conference on Data Engineering (ICDE)*. IEEE, 1033–1044.

[26] Xiao Huang, Jundong Li, and Xia Hu. 2017. Label informed attributed network embedding. In *Proceedings of the tenth ACM international conference on web search and data mining*. 731–739.

[27] Lawrence Hubert and Phipps Arabie. 1985. Comparing partitions. *Journal of classification* 2 (1985), 193–218.

[28] Pedro Jordano, Jordi Bascompte, and Jens M Olesen. 2003. Invariant properties in coevolutionary networks of plant–animal interactions. *Ecology letters* 6, 1 (2003), 69–81.

[29] Leonard Kaufman and Peter J Rousseeuw. 2009. *Finding groups in data: an introduction to cluster analysis*. Vol. 344. John Wiley & Sons.

[30] Brian W Kernighan and Shen Lin. 1970. An efficient heuristic procedure for partitioning graphs. *The Bell system technical journal* 49, 2 (1970), 291–307.

[31] Yuval Kluger, Ronen Basri, Joseph T Chang, and Mark Gerstein. 2003. Spectral biclustering of microarray data: coclustering genes and conditions. *Genome research* 13, 4 (2003), 703–716.

[32] Daniel B Larremore, Aaron Clauset, and Abigail Z Jacobs. 2014. Efficiently inferring community structure in bipartite networks. *Physical Review E* 90, 1 (2014), 012805.

[33] Sune Lehmann, Martin Schwartz, and Lars Kai Hansen. 2008. Biclique communities. *Physical review E* 78, 1 (2008), 016108.

[34] Yvonne Y Li and Steven JM Jones. 2012. Drug repositioning for personalized medicine. *Genome medicine* 4 (2012), 1–14.

[35] Xin Liu and Tsuyoshi Murata. 2010. Community detection in large-scale bipartite networks. *Transactions of the Japanese Society for Artificial Intelligence* 25, 1 (2010), 16–24.

[36] David Melamed. 2014. Community structures in bipartite networks: A dual-projection approach. *PloS one* 9, 5 (2014), e97823.

[37] Tsuyoshi Murata. 2009. Detecting communities from bipartite networks based on bipartite modularities. In *2009 International Conference on Computational Science and Engineering*, Vol. 4. IEEE, 50–57.

[38] Jose C Nacher and Jean-Marc Schwartz. 2012. Modularity in protein complex and drug interactions reveals new polypharmacological properties. *PloS one* 7, 1 (2012), e30028.

[39] Mark EJ Newman. 2006. Finding community structure in networks using the eigenvectors of matrices. *Physical review E* 74, 3 (2006), 036104.

[40] Mark EJ Newman and Michelle Girvan. 2004. Finding and evaluating community structure in networks. *Physical review E* 69, 2 (2004), 026113.

[41] Gergely Palla, Imre Derényi, Illés Farkas, and Tamás Vicsek. 2005. Uncovering the overlapping community structure of complex networks in nature and society. *nature* 435, 7043 (2005), 814–818.

[42] Paola Pesántez-Cabrera and Ananth Kalyanaraman. 2017. Efficient detection of communities in biological bipartite networks. *IEEE/ACM transactions on computational biology and bioinformatics* 16, 1 (2017), 258–271.

[43] Pascal Pons and Matthieu Latapy. 2005. Computing communities in large networks using random walks. In *International symposium on computer and information sciences*. Springer, 284–293.

[44] Alex Pothen, Horst D Simon, and Kang-Pu Liou. 1990. Partitioning sparse matrices with eigenvectors of graphs. *SIAM journal on matrix analysis and applications* 11, 3 (1990), 430–452.

[45] Benedek Rozemberczki and Rik Sarkar. 2020. Characteristic Functions on Graphs: Birds of a Feather, from Statistical Descriptors to Parametric Models. In *Proceedings of the 29th ACM International Conference on Information and Knowledge Management (CIKM '20)*. ACM, 1325–1334.

[46] Risa D Sargent and David D Ackerly. 2008. Plant–pollinator interactions and the assembly of plant communities. *Trends in Ecology & Evolution* 23, 3 (2008), 123–130.

[47] Prithviraj Sen, Galileo Namata, Mustafa Bilgic, Lise Getoor, Brian Galligher, and Tina Eliassi-Rad. 2008. Collective classification in network data. *AI magazine* 29, 3 (2008), 93–93.

[48] Arnab Sinha, Zhihong Shen, Yang Song, Hao Ma, Darrin Eide, and Kuansan Wang. 2015. An Overview of Microsoft Academic Service (MAS) and Applications. In *The WebConf*.

[49] Gilbert Strang, Gilbert Strang, Gilbert Strang, and Gilbert Strang. 1993. *Introduction to linear algebra*. Vol. 3. Wellesley-Cambridge Press.

[50] Alexander Strehl and Joydeep Ghosh. 2002. Cluster ensembles—a knowledge reuse framework for combining multiple partitions. *Journal of machine learning research* 3, Dec (2002), 583–617.

[51] Kenta Suzuki and Ken Wakita. 2009. Extracting multi-facet community structure from bipartite networks. In *2009 International Conference on Computational Science and Engineering*, Vol. 4. IEEE, 312–319.

[52] Raphael Tackx, Fabien Tarissan, and Jean-Loup Guillaume. 2018. ComSim: a bipartite community detection algorithm using cycle and node's similarity. In *Complex Networks & Their Applications VI: Proceedings of Complex Networks 2017 (The Sixth International Conference on Complex Networks and Their Applications)*. Springer, 278–289.

[53] Lei Tang and Huan Liu. 2009. Relational learning via latent social dimensions. In *Proceedings of the 15th ACM SIGKDD international conference on Knowledge discovery and data mining*. 817–826.

[54] Hanghang Tong, Christos Faloutsos, and Jia-Yu Pan. 2006. Fast random walk with restart and its applications. In *Sixth international conference on data mining (ICDM'06)*. IEEE, 613–622.

[55] Ulrike Von Luxburg. 2007. A tutorial on spectral clustering. *Statistics and computing* 17, 4 (2007), 395–416.

[56] Sibo Wang, Renchi Yang, Xiaokui Xiao, Zhewei Wei, and Yin Yang. 2017. FORA: simple and effective approximate single-source personalized pagerank. In *Proceedings of the 23rd ACM SIGKDD International Conference on Knowledge Discovery and Data Mining*. 505–514.

[57] David P Woodruff. 2014. Sketching as a Tool for Numerical Linear Algebra. (2014).

[58] Fangzhao Wu, Ying Qiao, Jiun-Hung Chen, Chuhan Wu, Tao Qi, Jianxun Lian, Danyang Liu, Xing Xie, Jianfeng Gao, Winnie Wu, and Ming Zhou. 2020. MIND: A Large-scale Dataset for News Recommendation. In *ACL*. 3597–3606.

[59] Yao Wu, Xudong Liu, Min Xie, Martin Ester, and Qing Yang. 2016. CCCF: Improving collaborative filtering via scalable user-item co-clustering. In *Proceedings of the ninth ACM international conference on web search and data mining*. 73–82.

[60] Rongkai Xia, Yan Pan, Lei Du, and Jian Yin. 2014. Robust multi-view spectral clustering via low-rank and sparse decomposition. In *Proceedings of the AAAI conference on artificial intelligence*, Vol. 28.

[61] Wei Xu, Xin Liu, and Yihong Gong. 2003. Document clustering based on non-negative matrix factorization. In *SIGIR*. 267–273.

[62] Renchi Yang. 2022. Efficient and Effective Similarity Search over Bipartite Graphs. In *Proceedings of the ACM Web Conference 2022*. 308–318.

[63] Renchi Yang, Jieming Shi, Keke Huang, and Xiaokui Xiao. 2022. Scalable and Effective Bipartite Network Embedding. In *Proceedings of the 2022 International Conference on Management of Data*. 1977–1991.

[64] Renchi Yang, Jieming Shi, Xiaokui Xiao, Yin Yang, and Sourav S Bhowmick. 2020. Homogeneous network embedding for massive graphs via reweighted

[65] Renchi Yang, Jieming Shi, Yin Yang, Keke Huang, Shiqi Zhang, and Xiaokui Xiao. 2021. Effective and scalable clustering on massive attributed graphs. In *Proceedings of the Web Conference 2021*. 3675–3687.

[66] Mao Ye, Dong Shou, Wang-Chien Lee, Peifeng Yin, and Krzysztof Janowicz. 2011. On the semantic annotation of places in location-based social networks. In *Proceedings of the 17th ACM SIGKDD international conference on Knowledge discovery and data mining*. 520–528.

[67] Tzu-Chi Yen and Daniel B Larremore. 2020. Community detection in bipartite networks with stochastic block models. *Physical Review E* 102, 3 (2020), 032309.

[68] Weihua Zhan, Zhongzhi Zhang, Jihong Guan, and Shuigeng Zhou. 2011. Evolutionary method for finding communities in bipartite networks. *Physical review E* 83, 6 (2011), 066120.

[69] Tian Zhang, Raghu Ramakrishnan, and Miron Livny. 1996. BIRCH: an efficient data clustering method for very large databases. *SIGMOD* (1996), 103–114.

[70] Tao Zhou, Jie Ren, Matúš Medo, and Yi-Cheng Zhang. 2007. Bipartite network projection and personal recommendation. *Physical review E* 76, 4 (2007), 046115.

[71] Katharina Anna Zweig and Michael Kaufmann. 2011. A systematic approach to the one-mode projection of bipartite graphs. *Social Network Analysis and Mining* 1 (2011), 187–218.

personalized PageRank. *Proceedings of the VLDB Endowment* 13, 5 (2020), 670–683.